\documentclass[aps,pra,reprint,groupedaddress]{revtex4-1}

\usepackage[]{graphicx}	
\usepackage{stackengine}
\usepackage{subfig}
\usepackage{amsmath}
\usepackage{amsfonts}

\begin{document}


\title{Annealing Dynamics via Quantum Interference of\\Forward and Backward Time Evolved States}

\author{Kentaro IMAFUKU}
\affiliation{National Institute of Advanced Industrial Science and Technology (AIST)}

\date{\today}

\begin{abstract}
Toward an alternative approach to the quantum mechanic ground state search, we theoretically introduce a protocol in which  energy of two identical systems are deterministically transferred. The protocol utilizes a quantum interference between ``forward" and ``backward" time evolved states with respect to a given Hamiltonian. In addition, to make use the protocol for the ground state search, we construct a network with which we may be able to efficiently apply the protocol successively among multiple systems so that energy of one of them is gradually approaching the lowest one. Although rigorous analysis on the validity of the network is left as a future challenge, some properties of the network are also investigated.
\end{abstract}


\maketitle


\section{Introduction\label{sec:1}}
A beautiful quantum extension of annealing computation -- so-called quantum annealing computation -- has been proposed \cite{PhysRevB.39.11828,PhysRevE.58.5355,2000quant.ph..1106F,Farhi2000ANS,PhysRevA.65.042308,RevModPhys.80.1061,2011Dwavenature,PhysRevX.4.021041,2014McGeoch}, and  is paid lots of attentions today even from industries. Similarly to the idea of the classical annealing computation\cite{Kirkpatrick671}, a problem to be addressed is mapped into a so-called problem Hamiltonian $\hat{H}$ whose ground state corresponds to the solution of the problem. In the quantum annealing computation, on the other hand, the ground state search is quantum mechanically carried out by driving the physical state according to a quantum dynamics. In particular, a specific property of the adiabatic process in quantum dynamics is fully utilized there. Introducing a driving Hamiltonian $\hat{V}$ which is non-commutative with $\hat{H}$, a time dependent Hamiltonian such as
\begin{equation}
\hat{H}(t)=s(t)\hat{H}+ \left(1-s(t)\right)\hat{V}
\end{equation}
is considered with a controlling function $s(t)\in {\mathbb R}$ that typically behaves monotonically as $s(0)=0$ and $s(t_f)=1$. ($t_f$ is the final time of the annealing process.) According to the adiabatic theorem in the quantum mechanics\cite{Born1928,doi:10.1143/JPSJ.5.435,messiah1999quantum}, when the state of the system is prepared in the ground state of $\hat{V}$ initially ($t=0$), the state evolved by the time dependent Hamiltonian is approximately maintained to be the ground state of instantaneous Hamiltonian $\hat{H}(t)$ at the each moment, {\it if the time dependence of $s(t)$ is sufficiently gentle}.  When the condition is satisfied, as the consequence of the adiabatic theorem, we can find the ground state of $\hat{H}$ with a high probability in a measurement to be performed at $t=t_f$. In the evolution, the state can become highly non-classical and can quantum mechanically shortcut the classical path of computation (or quantum tunneling) if necessary.  Thus, the computation is expected to be superior to the classical annealing computation. 

On the other hand, the speed of the quantum annealing computation is restrictively determined by the gap ($:=g_{min}(t)$) between the energy levels of the ground state ($|g(t)\rangle$) and the first excited state ($|e(t)\rangle$) of the instantaneous Hamiltonian. More specifically, 
$$
\frac{\max_{t\in [0,t_f]} |\langle e(t)|\frac{d}{dt} \hat{H}(t)|g(t)\rangle|}{\min_{t\in[0,t_f]} g_{min}(t)^2}\ll 1
$$
is required to achieve the appropriate adiabatic process (and that is the exact reason that the gentleness of $s(t)$ is required.) In other words, when there exists a small gap, the time derivative of the control function cannot be large so much, and $t_f$ becomes unavoidably large. Unfortunately, some examples indicating that a quantum first order phase transition tends to occur during the adiabatic computation \cite{0305-4470-39-36-R01,Santoro2427,PhysRevA.80.062326,PhysRevLett.104.020502,PhysRevLett.104.207206,PhysRevE.84.061152,ncomms12370} have been found. In these cases, $\min_{t\in[0,t_f]}g_{min}(t)$ becomes exponentially small with respect to the size of the system, and implies an exponential slowing down of the speed of the computation. Although various interesting investigations to avoid the phase transition by appropriate choices of $\hat{V}$ and $s(t)$ are being tried \cite{PhysRevE.85.051112,1751-8121-48-33-335301,10.3389/fict.2017.00002,doi:10.7566/JPSJ.87.023002}, the slowing down can be a fundamental bottleneck of the existing quantum annealing computation.

With the circumstances, we propose an alternative approach to the quantum mechanic ground state search. The approach consists of the following two parts:  (1)Introduction of an energy transfer protocol between two systems, and (2)Network structure to efficiently apply the protocol  to the ground state search. Combining the two ideas, we aim to gradually remove the energy of a system so as the state of the system efficiently achieves  the ground state of the Hamiltonian. 

We describe our idea as follows: In the next section, we introduce the energy transfer protocol between two systems. The protocol is determined only by the problem Hamiltonian. There, we will find that the protocol interestingly utilizes a quantum interference between “forward” and “backward” time evolved states in terms of the Hamiltonian. In SEC.\ref{sec:3}, we show that the energy transfer by the above protocol can be described in a short time behavior of the solution of a certain nonlinear Schr\"{o}dinger equation.  In addition, a property of the solution efficiently converging to the ground state of the problem Hamiltonian is demonstrated. In SEC.\ref{sec:4} , aiming a physical emulation of the nonlinear Schr\"{o}dinger equation beyond the short time behavior, we propose a network with which we can apply the protocol successively among multiple systems. Although this part remains further challenges that should be carefully clarified, some analysis on the validity of the network are also discussed in SEC.\ref{sec:5}.

\section{Energy Transfer Protocol\label{sec:2}}
In this section, we introduce an energy transfer protocol between two systems. Suppose that we have two systems ({\bf a} and {\bf b}) which are in a same state $|\varphi_0\rangle$. (Total state of the two system is $|\varphi_0\rangle\otimes|\varphi_0\rangle$ in Hilbert space ${\mathcal H}_a\otimes{\mathcal H}_b\simeq{\mathcal H}^{\otimes 2}$.)
With respect to Hamiltonian $\hat{H}$ on each Hilbert space, both systems have the same expectation value, i.e.,
\begin{equation}
E_0:=\langle\varphi_0|\hat{H}|\varphi_0\rangle.
\end{equation}
In the following, we introduce a protocol among the two systems that realizes
\begin{equation}\label{eq:2}
{\rm Tr}\left(\rho_a \hat{H}\right) \le E_0 \le {\rm Tr}\left(\rho_b \hat{H}\right)
\end{equation}
where $\rho_a$ and $\rho_b$ are the reduced states of systems {\bf a} and {\bf b} obtained by the protocol respectively.
The protocol works independently from the specific forms of Hamiltonian and the state $|\varphi_0\rangle$. 
\bigskip

\noindent{\bf Protocol:}
\begin{enumerate}
\item Prepare a initial state
\begin{equation}
|\Psi_{in}\rangle:=|\varphi_0\rangle \otimes|\varphi_0\rangle
\end{equation}
in Hilbert space ${\mathcal H}_a\otimes{\mathcal H}_b$. 
\item Acting unitary operation described as 
\begin{equation}\label{eq:5}
U=e^{+i \hat{S}_{ab} ~\pi/4} \left(e^{-i\hat{H}t/2}\otimes e^{+i\hat{H}t/2}\right)
\end{equation}
where $\hat{S}_{ab}$ is a swapping operator among ${\mathcal H}_a\otimes {\mathcal H}_b(\simeq {\mathcal H}^{\otimes 2})$ that holds
\begin{equation}
\hat{S}_{ab} |j\rangle\otimes |k\rangle=|k\rangle\otimes|j\rangle
\end{equation}
with a given orthonormal basis $\{|j\rangle\}$ for each Hilbert space ${\mathcal H}$.
\end{enumerate}
(Similar idea of the usage of the swapping operation for making quantum mechanic time evolution of one system depend on another quantum state can be found in \cite{Lloyd2014}.)

\begin{figure*}[t]
\centering
        \subfloat[Idea of protocol]{
            \includegraphics[scale=0.4]{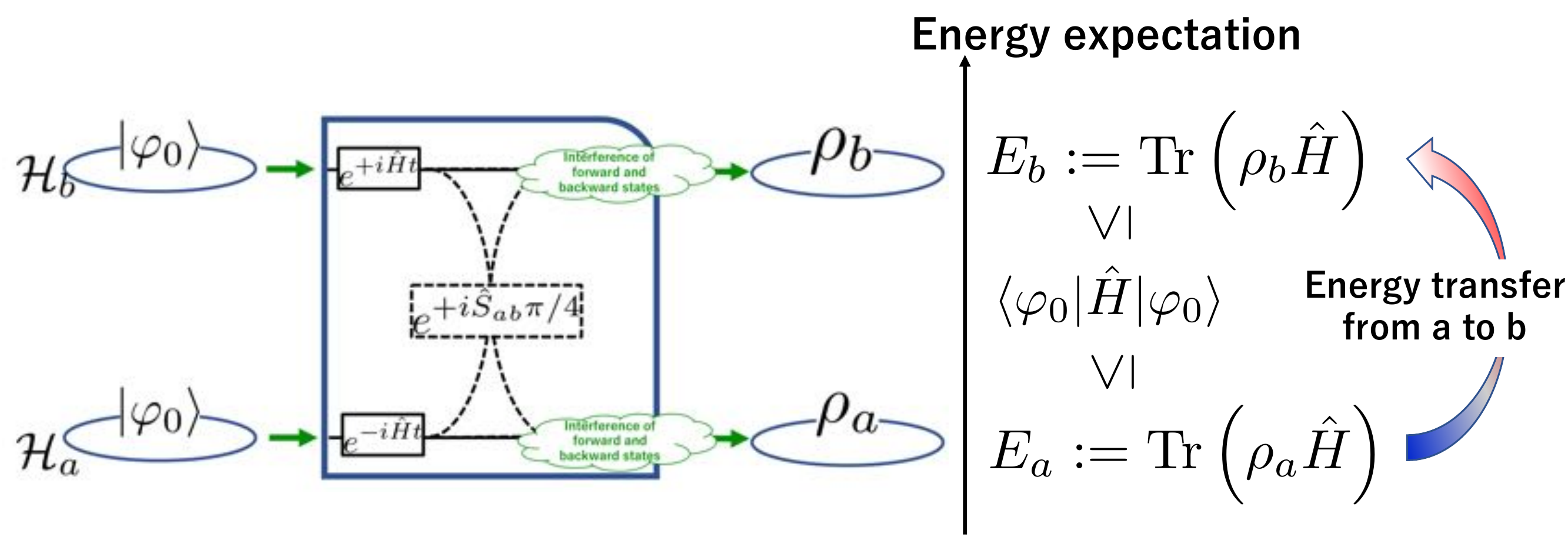}
        }~~~~~
         \subfloat[Simple notation]{
            \includegraphics[scale=0.2]{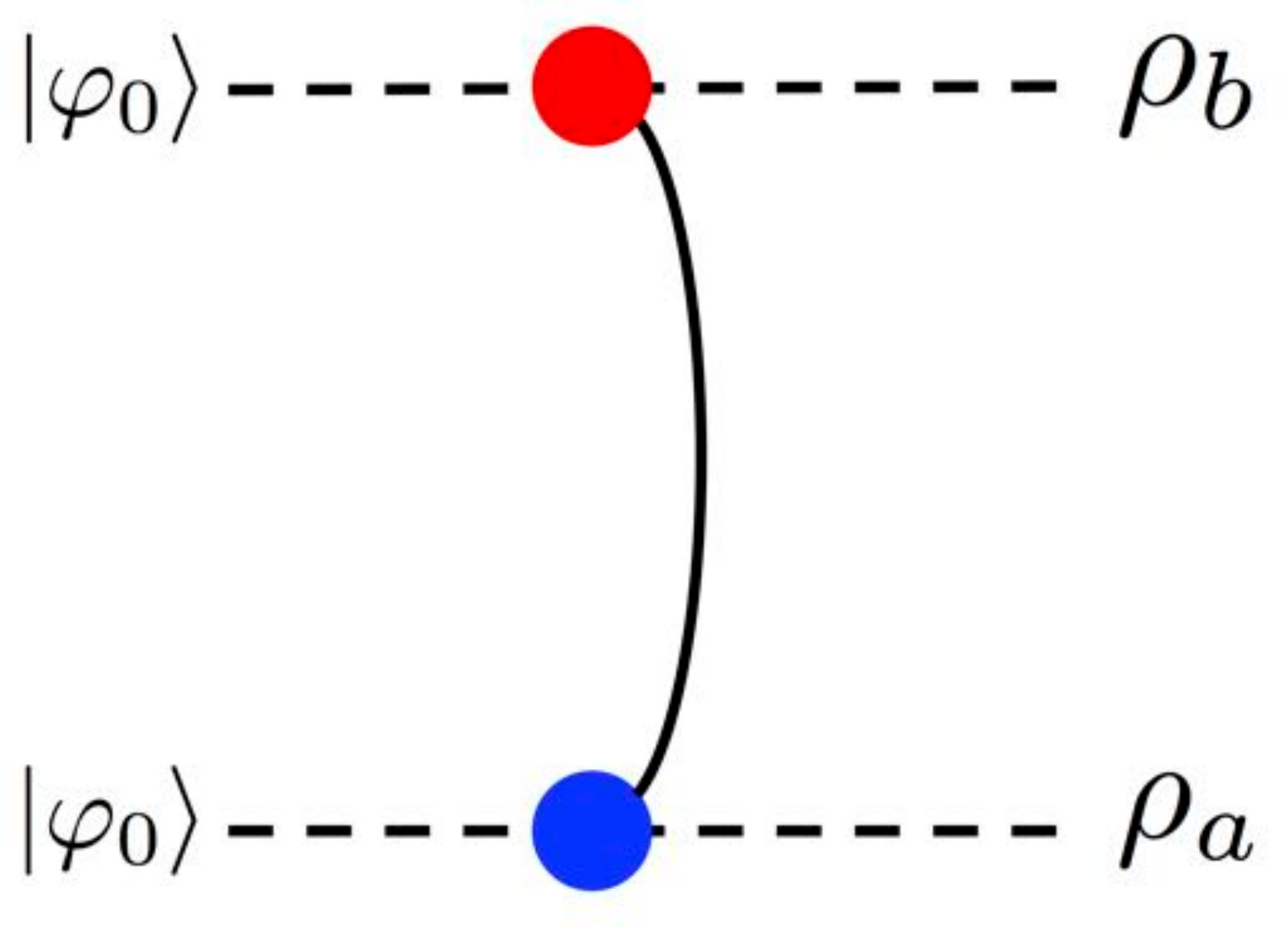}
        }
        \caption{{\bf Schematic Illustration of the energy transfer protocol.} In FIGs.\ref{fig:fig3} and \ref{fig:4}, the protocol is represented by the simple notation in (b).} 
\end{figure*}
Notice that the unitary operation in (\ref{eq:5}) contain a swapping operation between ``forward evolved state $e^{-i\hat{H} t/2} |\varphi_0\rangle$" and ``backward evolved state $e^{+i\hat{H} t/2} |\varphi_0\rangle$". Following the above protocol, we obtain
\begin{eqnarray}
\rho_a&&:={\rm Tr}_{b}\Big(U|\Psi_{in}\rangle\langle\Psi_{in}|U^\dagger\Big)\nonumber\\
=&&\frac{1}{2} \left(e^{-i\hat{H}t/2} |\varphi_0\rangle\langle\varphi_0| e^{+i\hat{H}t/2}+e^{+i\hat{H}t/2} |\varphi_0\rangle\langle\varphi_0| e^{-i\hat{H}t/2}\right)\nonumber\\
&&+i\frac{1}{2}\langle\varphi_0|e^{-i\hat{H}t}|\varphi_0\rangle ~e^{+i\hat{H}t/2}|\varphi_0\rangle\langle\varphi_0|e^{+i\hat{H}t/2}\nonumber\\
&&-i\frac{1}{2}\langle\varphi_0|e^{+i\hat{H}t}|\varphi_0\rangle ~e^{-i\hat{H}t/2}|\varphi_0\rangle\langle\varphi_0|e^{-i\hat{H}t/2}\label{eq:8}
\end{eqnarray}
and
\begin{eqnarray}
\rho_b&&:={\rm Tr}_{a}\Big(U|\Psi_{in}\rangle\langle\Psi_{in}|U^\dagger\Big)\nonumber\\
=&&\frac{1}{2} \left(e^{-i\hat{H}t/2} |\varphi_0\rangle\langle\varphi_0| e^{+i\hat{H}t/2}+e^{+i\hat{H}t/2} |\varphi_0\rangle\langle\varphi_0| e^{-i\hat{H}t/2}\right)\nonumber\\
&&-i\frac{1}{2}\langle\varphi_0|e^{-i\hat{H}t}|\varphi_0\rangle ~e^{+i\hat{H}t/2}|\varphi_0\rangle\langle\varphi_0|e^{+i\hat{H}t/2}\nonumber\\
&&+i\frac{1}{2}\langle\varphi_0|e^{+i\hat{H}t}|\varphi_0\rangle ~e^{-i\hat{H}t/2}|\varphi_0\rangle\langle\varphi_0|e^{-i\hat{H}t/2}\label{eq:9}
\end{eqnarray}
where ${\rm Tr}_{b}$ and ${\rm Tr}_{a}$ are partial trace operations over ${\mathcal H}_b$ and ${\mathcal H}_a$ respectively. The second and third lines in (\ref{eq:8}) and (\ref{eq:9}) are interference terms of the forward and backward evolved states.  Notice that, because of the existence of the interference terms, an energy transfer among the two systems occurs. In fact, we can find that
\begin{equation}
E_a:={\rm Tr}\left(\rho_a\hat{H}\right)=E_0 +\frac{1}{2}\frac{d}{dt} P_0(t)\label{eq:10}
\end{equation}
and
\begin{equation}
E_b:={\rm Tr}\left(\rho_b\hat{H}\right)=E_0 -\frac{1}{2}\frac{d}{dt} P_0(t)\label{eq:11}
\end{equation}
hold where $P_0(t)$ is the survival probability defined as
\begin{equation}
P_0(t)=|\langle \varphi_0|e^{-i\hat{H}t}|\varphi_0\rangle|^2.
\end{equation}
Considering small $t (=\delta t>0)$, we can estimate (\ref{eq:10}) and (\ref{eq:11}) as
\begin{equation}
E_a=E_0- \delta E^2_0 \delta t+O(\delta t^3)\label{eq:13}
\end{equation}
and
\begin{equation}
E_b=E_0+\delta E^2_0 \delta t+O(\delta t^3)\label{eq:14}
\end{equation}
where $\delta E^2_0 >0$ is the energy variance defined as
\begin{equation}
\delta E^2_0:=\langle\varphi_0|\hat{H}^2|\varphi_0\rangle-\langle\varphi_0|\hat{H}|\varphi_0\rangle^2. 
\end{equation}
With a small enough $\delta t$ to safely ignore $O(\delta t^3)$, the above indicates an energy transfer proportional to the variance from system {\bf a} to {\bf b}.

\section{Non-Linear Schr\"{o}dinger equation\label{sec:3}}
In this section, we show that there exists a non-linear Schr\"{o}dinger equation approximately corresponding to the protocol introduced in the previous section.
Expanding (\ref{eq:8}) and (\ref{eq:9}) by $t (=\delta t >0)$, we obtain
\begin{eqnarray}
\label{eq:16}
\!\!
\rho_a&&=|\varphi_0\rangle\langle\varphi_0|-i\delta t \left[\hat{G}_0,|\varphi_0\rangle\langle\varphi_0|\right]\nonumber\\
&&\!\!\!
-\frac{\delta t^2}{8}\left(\left\{\hat{H}^2,|\varphi_0\rangle\langle\varphi_0|\right\}-\hat{H}|\varphi_0\rangle\langle\varphi_0|\hat{H}\right)+O(\delta t^3)
\\
\label{eq:17}
\!\!
\rho_b&&= |\varphi_0\rangle\langle\varphi_0|+i\delta t \left[\hat{G}_0,|\varphi_0\rangle\langle\varphi_0|\right]\nonumber\\
&&\!\!\!
-\frac{\delta t^2}{8}\left(\left\{\hat{H}^2,|\varphi_0\rangle\langle\varphi_0|\right\}-\hat{H}|\varphi_0\rangle\langle\varphi_0|\hat{H}\right)+O(\delta t^3)
\end{eqnarray}
where $\hat{G}_0$ is an hermitian operator defined as
\begin{equation}\label{eq:18}
\hat{G}_0:=-i\left[\frac{\hat{H}}{2},|\varphi_0\rangle\langle\varphi_0|\right].
\end{equation}
Generalizing (\ref{eq:16}), (\ref{eq:17}) and (\ref{eq:18}), let us introduce the following non-linear Schr\"{o}dinger equation:
\begin{equation}\label{eq:19}
\frac{d}{dt}|\varphi_t\rangle=-i\hat{G}_t |\varphi_t\rangle
\end{equation}
with a state-dependent hermitian operator
\begin{equation}\label{eq:20}
\hat{G}_t=-i\left[\frac{\hat{H}}{2},|\varphi_t\rangle\langle\varphi_t|\right].
\end{equation}
(Interestingly, in literatures\cite{Huignard:81,Ja1982,Gunter1982199,PhysRevLett.82.1418} on the two beams coupling phenomena in photorefractive media, we find that equations with similar structure to (\ref{eq:19}) and (\ref{eq:20}) have been phenomenologically introduced.)
Employing an initial state as 
\begin{equation}\label{eq:21}
|\varphi_{t=0}\rangle=|\varphi_0\rangle,
\end{equation}
we obtain
\begin{eqnarray}
\label{eq:22}\rho_a&&=|\varphi_{+\delta t}\rangle\langle\varphi_{+\delta t}|\nonumber\\
&&-\delta t^2\Big(\left\{\hat{\Gamma}_0, |\varphi_{t=0}\rangle\langle\varphi_{t=0}|\right\}-2 \langle\varphi_{t=0}|\hat{\Gamma}_0|\varphi_{t=0}\rangle\Big)
\nonumber\\
&&+O(\delta t^3)\\
\label{eq:23}\rho_b&&=|\varphi_{-\delta t}\rangle\langle\varphi_{-\delta t}|\nonumber
\\
&&-\delta t^2\Big(\left\{\hat{\Gamma}_0, |\varphi_{t=0}\rangle\langle\varphi_{t=0}|\right\}-2 \langle\varphi_{t=0}|\hat{\Gamma}_0|\varphi_{t=0}\rangle\Big)\nonumber\\
&&+O(\delta t^3)
\end{eqnarray}
with
\begin{equation}\label{eq:24}
\hat{\Gamma}_0=\frac{1}{4}\left(\hat{H}-\langle\varphi_{t=0}|\hat{H}|\varphi_{t=0}\rangle\right)^2.
\end{equation}
Equations (\ref{eq:22}) and (\ref{eq:23}) suggest that $\rho_a$ and $\rho_b$ obtained by the above protocol approximately emulates the short time dynamics described in the non-linear Schr\"{o}dinger equation in (\ref{eq:19}) in forward ($+\delta t$) and backward ($-\delta t$) directions respectively. We utilize the fact to propose an efficient combination of the protocol among systems to  physically emulate the non-linear dynamics for finite time interval (beyond the short time dynamics). Before the proposal, let us remark some properties on the non-linear Schr\"{o}dinger equation itself.

\begin{figure*}[t]
\centering
        \subfloat[]{
            \includegraphics[scale=0.4]{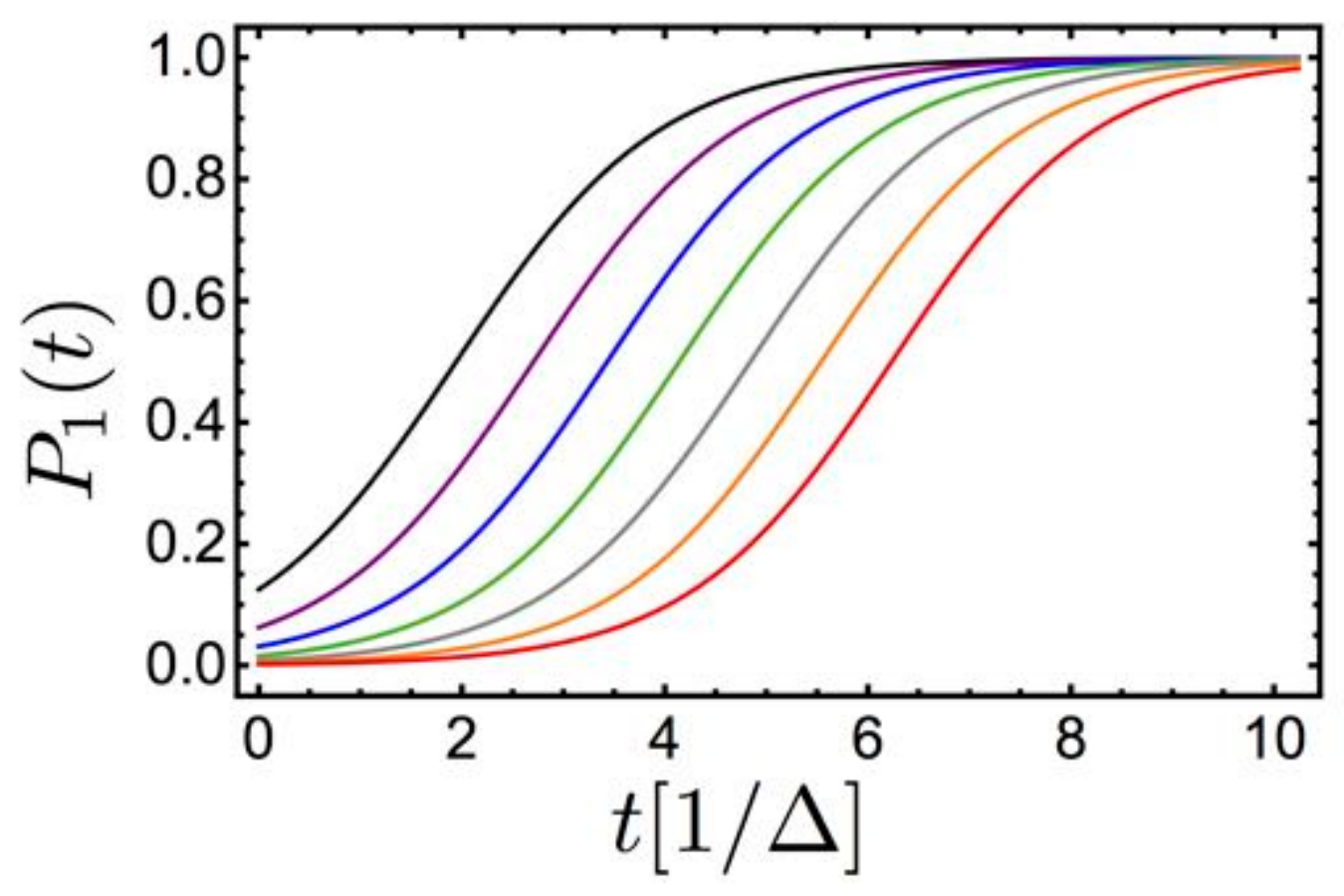}
        }~~~~~~~~~~~~
        \subfloat[]{
            \includegraphics[scale=0.4]{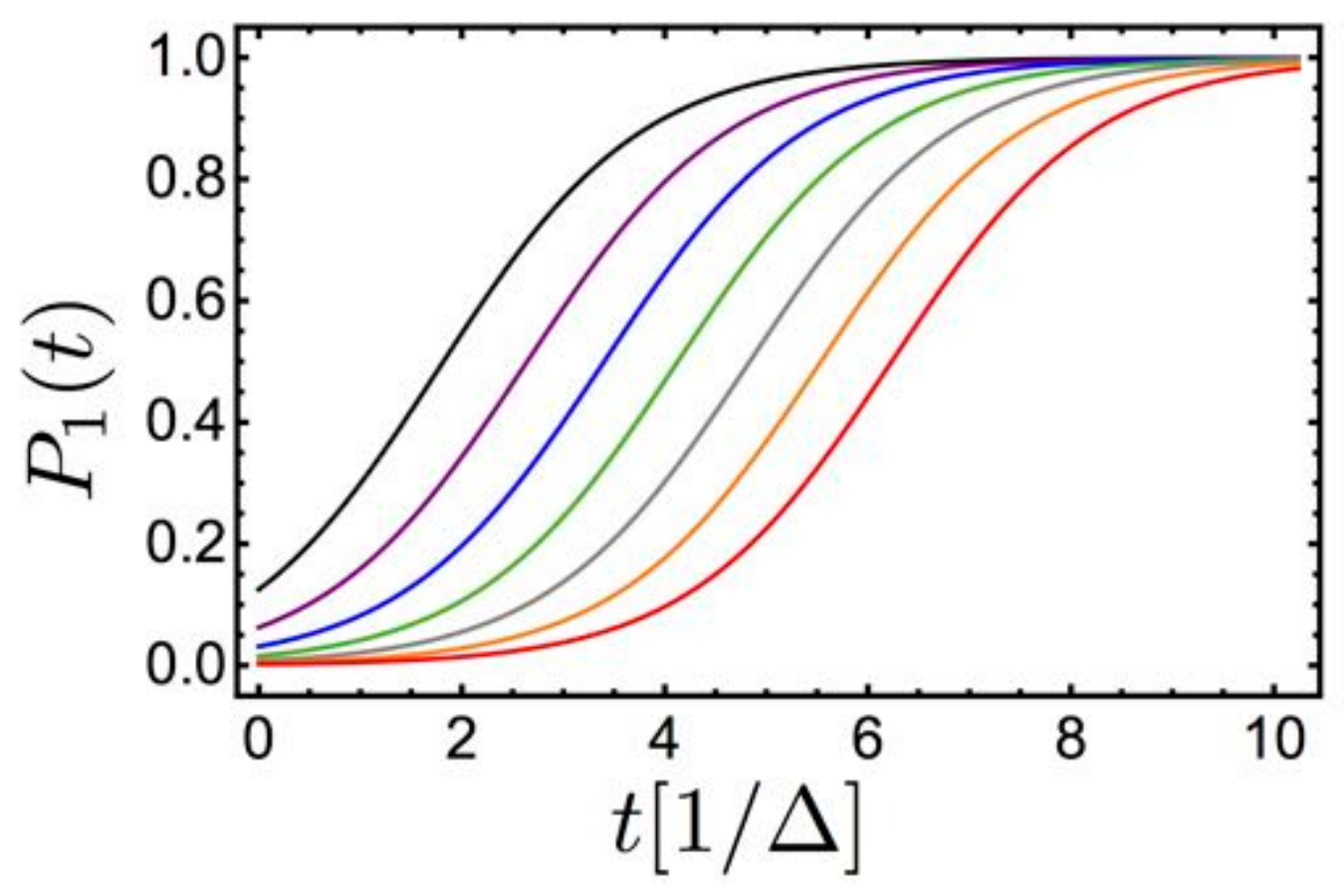}
        }\\
         \subfloat[]{
            \includegraphics[scale=0.4]{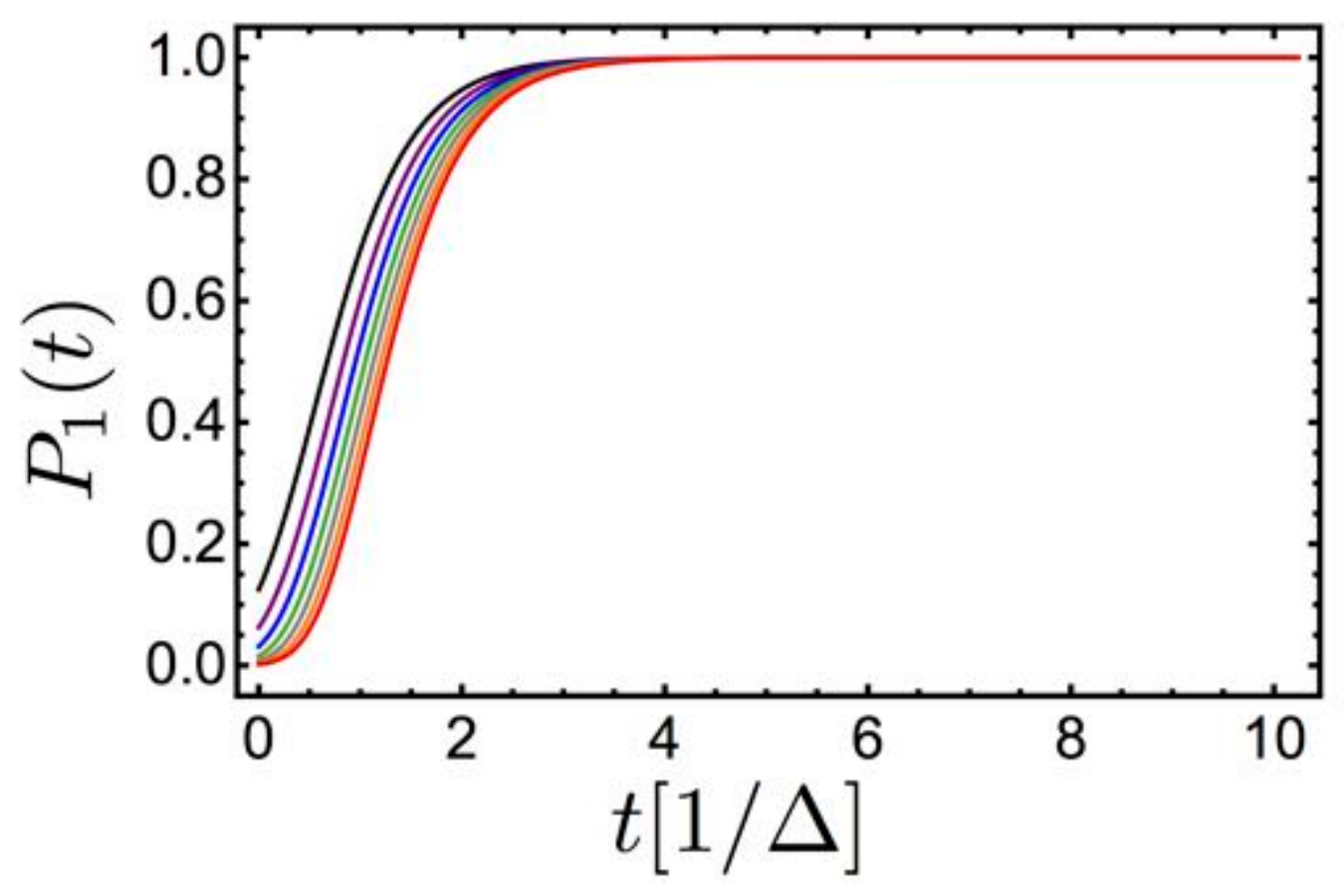}
        }~~~~~~~~~~~~
        \subfloat[]{
            \includegraphics[scale=0.4]{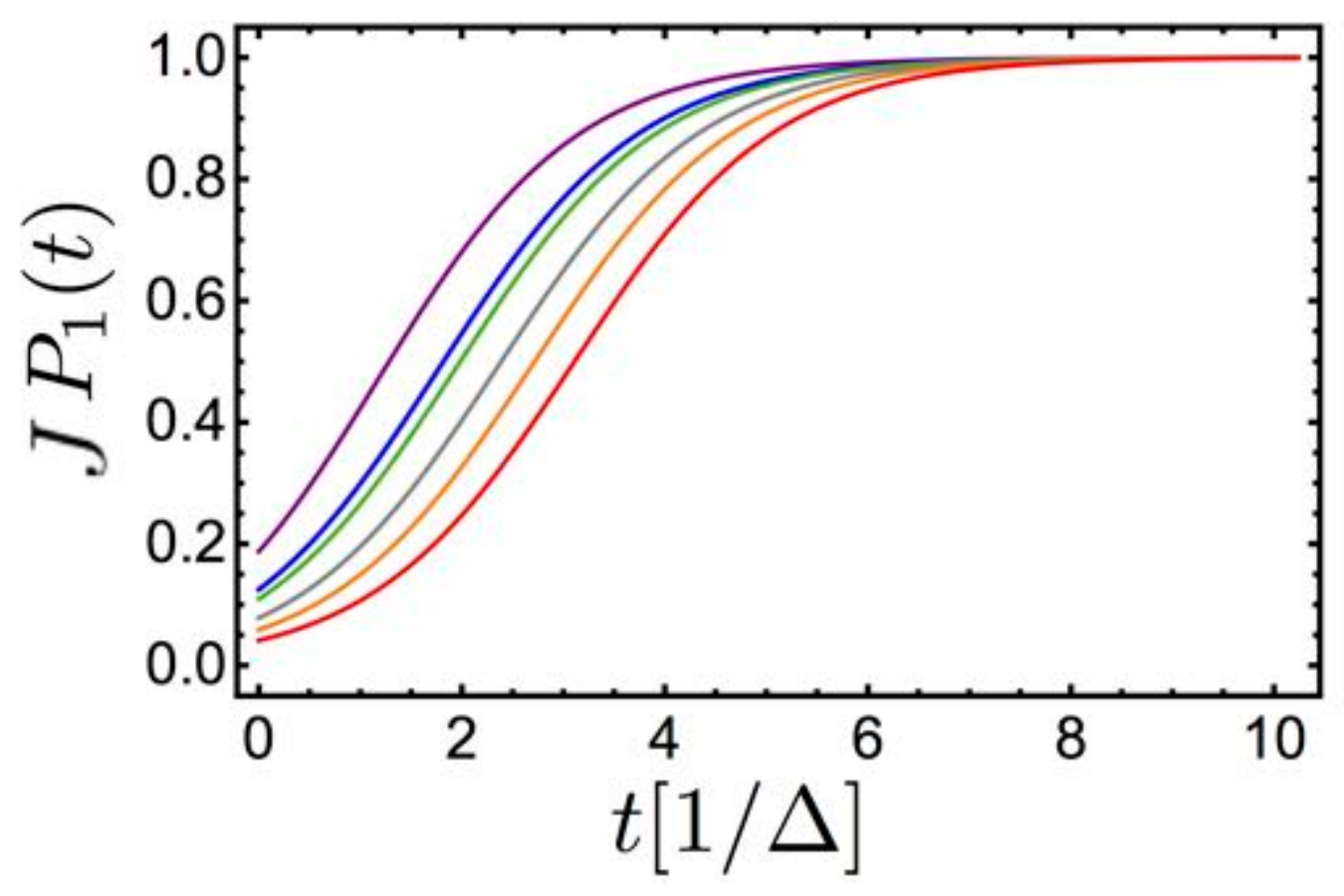}
        }
\caption{{\bf Hamiltonian dependence of $P_1(t)$.} Four Hamiltonians defined in TABLE \ref{table:table1} are exampled. In each panel, each line from top to bottom corresponds to the case of  $\dim({\mathcal H})=2^3$ (black), $2^4$ (purple), $2^5$ (blue), $2^6$ (green), $2^7$ (magenta), $2^8$ (orange), and $2^{9}$ (red), respectively. In all cases, the initial state is chosen according to (\ref{eq:27}). (In (d), degree of degeneracy $J$ is multiplied to $P_1(t)$. See TABLE \ref{table:table1} for the definition of $J$.)
\label{fig:fig2}}
\end{figure*}

In the following, let us suppose a form of Hamiltonian as
\begin{equation}\label{eq:26}
\hat{H}=\sum_{j=1}^{\dim({\mathcal H})} \varepsilon_j~ |\varepsilon_j\rangle\langle \varepsilon_j|
\end{equation}
with
\begin{equation}
\varepsilon_1 < \varepsilon_2 \le\cdots \le \varepsilon_{\dim({\mathcal H})}.
\end{equation}
(For simplicity, we assume that there is no degeneracy in its ground state.)
Remember that $\dim({\mathcal H})$ is the size of the search space. Unless $\langle\varepsilon_1|\varphi_{t=0}\rangle$ is exactly $0$,
the probability $P_1(t):=|\langle\varepsilon_1|\varphi_t\rangle|^2$ grows exponentially and converges to one. 
(Remember that $|\varphi_t\rangle$ is the solution of eq.(\ref{eq:19}) with eq.(\ref{eq:20}).)
For example, when we chose initial state $|\varphi_{t=0}\rangle$ as
\begin{equation}\label{eq:27}
|\varphi_{t=0}\rangle=\frac{1}{\sqrt{\dim({\mathcal H})}}\sum_{j=1}^{\dim({\mathcal H})}|\varepsilon_j\rangle,
\end{equation}
we can proof that
\begin{equation}
\frac{1}{\left(\dim({\mathcal H})-1\right) \exp(-D t)+1}\le P_1(t) 
\end{equation}
and
\begin{equation}
P_1(t) \le \frac{1}{\left(\dim({\mathcal H})-1\right) \exp(-\Delta t)+1}
\end{equation}
hold, where
\begin{equation}
\Delta:=\varepsilon_2-\varepsilon_1,\quad\mbox{and}\quad D:=\varepsilon_{\dim({\mathcal H})}-\varepsilon_1.
\end{equation}
(See appendix for the derivation of the inequality.) By the inequality, for $t_c$ such as $P_1(t_c)=c$, we find
\begin{equation}\label{eq:30}
\frac{1}{D}\log \frac{\dim({\mathcal H})-1}{1/c-1}
\le t_c \le
\frac{1}{\Delta}\log \frac{\dim({\mathcal H})-1}{1/c-1}.
\end{equation}
From the computational complexity theoretic point of view, $c$ is not necessarily to be a constant. In general
\begin{equation}
1/\mbox{poly of $\log(\dim ({\mathcal H}))$}
\end{equation}
is sufficient as an interesting choice of $c$. (Remember that $\dim{\mathcal H}$ is corresponding to the size of the search space.) Notice that $\Delta$ --the gap of the problem Hamiltonian itself-- does not depend on $\dim({\mathcal H})$ in a reasonable setting, whereas $g_{min}(t)$ in the quantum annealing process are apt to do (See Sec.\ref{sec:1}). The dependence in the latter approach implies time scales for the annealing process to be polynomial in $\dim({\mathcal H})$ that is exponential to $t_c$ in (\ref{eq:30}).  
In fact, in FIG.\ref{fig:fig2}, efficient convergences to $1$ of $P_1(t)$ are numerically shown with respect to some examples.
They indicate that {\it if we can efficiently emulate the dynamics of the non-linear Schr\"{o}dinger equation up to the time scale $t_c$}, we may be able to utilize the emulation to find the ground state of a given Hamiltonian more efficiently than by the adiabatic annealing process. That motivates us to extend the protocol in the previous section (that corresponds to the short time emulation of the non-linear Schr\"{o}dinger equation) to achieve the finite time emulation.

\begin{table*}[t]
 \begin{tabular}{c|c|l}
\hline\hline
Hamiltonian & Implication & Definition of Hamiltonian with (\ref{eq:26})\\ \hline
    (a) &Database search & 
\begin{tabular}{ll}
    $\varepsilon_j=-\Delta$ & for $j=1$, and\\
    $\varepsilon_j=0$ & for others.
\end{tabular}    
    \\
    \hline
    (b) & A symmetrization of (a) & 
    \begin{tabular}{ll}
    $\varepsilon_j=-\Delta$ & for $j=1$,\\
    $\varepsilon_j=0$ & for $1<j<\dim({\mathcal H})$, and\\
    $\varepsilon_j=+\Delta$ & for $j=\dim({\mathcal H})$.
\end{tabular}   
\\
    \hline
    (c) & Binomial distribution &
\begin{tabular}{l}
$\varepsilon_j=-\Delta \left(B_0\left[j-1,\dim({\mathcal H})\right]-B_1\left[j-1,\dim({\mathcal H})\right]\right)$,\\
where $B_q[x,y]$ is the number of $q$ in the binary \\ representation of $x$ in $\log_2 y$-digits.
\end{tabular}
 \\
    \hline
    (d) &Another symmetrization of (a)&
\begin{tabular}{ll}   
    $\epsilon_j=-\Delta$ & for $1\le j\le J$,\\
    $\epsilon_j=0$ & for $j_0< j <\dim(\mathcal H)-J$, and\\
    $\epsilon_j=+\Delta$ & for $\dim(\mathcal H)-J \le j \le\dim({\mathcal H})$\\
    & with $J=\mbox{Integer part of}~\sqrt{\dim({\mathcal H})}-1$.
\end{tabular}\\
    \hline\hline
  \end{tabular}
  \caption{\label{table:table1}{\bf Hamiltonians examined in FIGs.\ref{fig:fig2} and \ref{fig:5}.}}
\end{table*}

\begin{figure}[t]
\begin{center}
\includegraphics[scale=0.3]{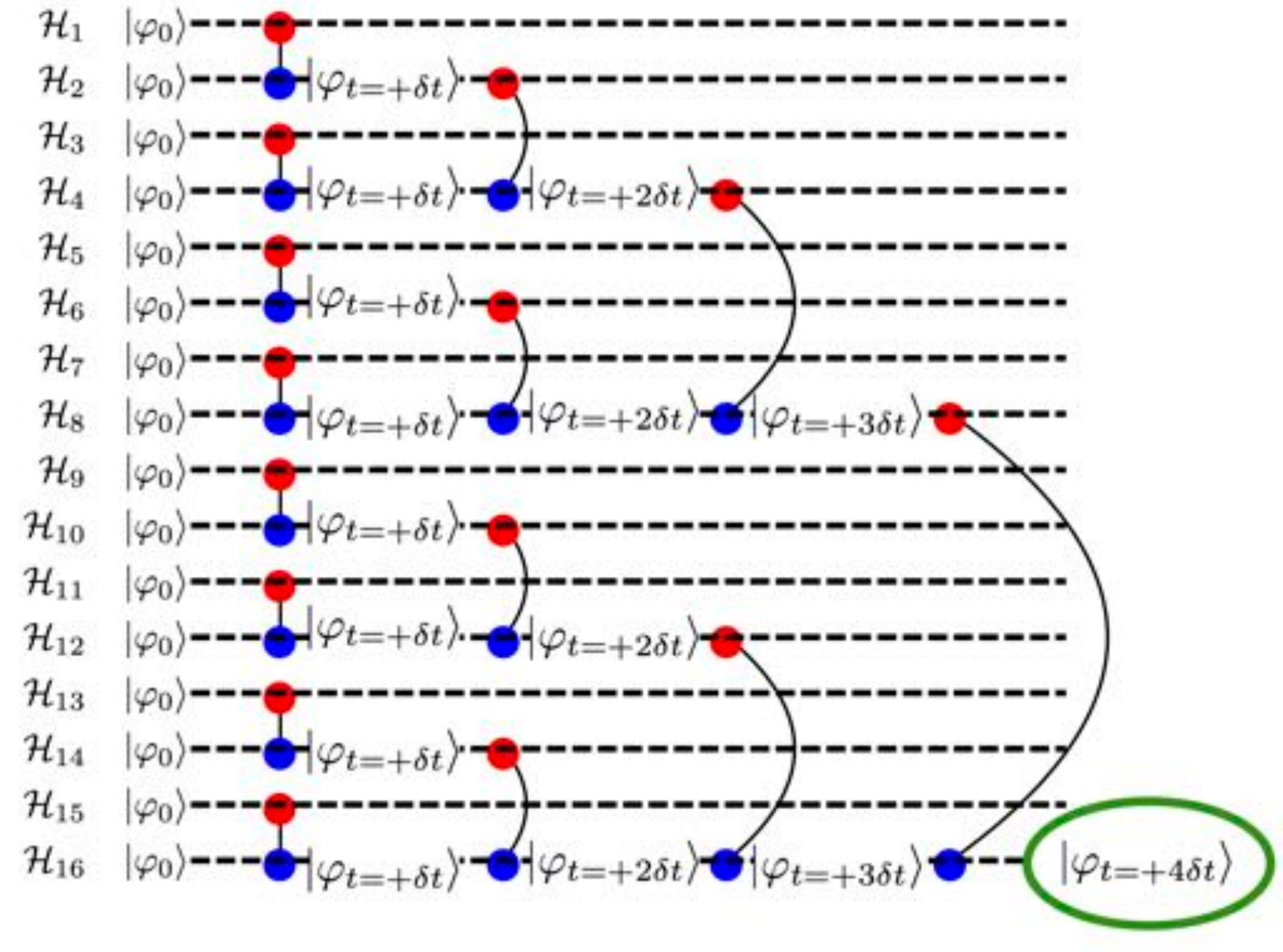}
\end{center}
\caption{{\bf Tournament Network.} The network requires  $2^4=16$ systems to execute the four steps achieving $|\varphi_{+4\delta t}\rangle$.
\label{fig:fig3}}
\end{figure}

Notice that, if we disregard efficiency in terms of the number of systems, there exists a trivial way of such extension. Applying the energy transfer protocol in a `tournament manner' as is shown in FIG.\ref{fig:fig3}, we approximately obtain $|\varphi_{t=+n\delta t}\rangle$ using $2^n$ systems.
More precisely, the final state in Hilbert space ${\mathcal H}_{2^n}$ becomes
\begin{eqnarray}
\rho_{2^{n}}(+n\delta t)&&=|\varphi_{t=+n\delta t}\rangle\langle\varphi_{t=+n\delta t}|\nonumber\\
&&\!\!\!\!\!\!\!\!\!\!\!\!\!\!\!\!\!\!\!\!\!\!
-\delta t^2
\sum_{k=0}^{n-1}\left(\{\hat{\Gamma}_k,|\varphi_{t=+k\delta t}\rangle\langle\varphi_{t=+k\delta t}|\}\right.\nonumber\\
&&\left.-2\langle\varphi_{t=+k\delta t}|\hat{\Gamma}_{k}|\varphi_{t=+k\delta t}\rangle\right)+O(\delta t^3)
\label{eq:31}
\end{eqnarray}
where $\hat{\Gamma}_k$ is defined similarly to (\ref{eq:24}) as
\begin{equation}
\hat{\Gamma}_k=\frac{1}{4}\left(\hat{H}-\langle\varphi_{t=+k\delta t}|\hat{H}|\varphi_{t=+k\delta t}\rangle\right)^2.
\end{equation}
Putting $n=t_c/\delta t$, we obtain
\begin{equation}\label{eq:33}
\langle\varphi_{t=t_c}| \rho_{2^{n}}(t_c)|\varphi_{t=t_c}\rangle>1-O(\delta t).
\end{equation}
In the tournament way, however, $2^{n}$ systems will be required to achieve the emulation up to $t_c=n\delta t$. In addition, to make the dynamics in (\ref{eq:20}) safely approximated by the discretized time steps, $\delta t < D^{-1}$ is also required. Combining the requirement for $\delta t$ with (\ref{eq:30}), the required number of the systems is estimated as
\begin{equation}\label{eq:34-1}
2^n\sim O(\dim({\mathcal H}))
\end{equation}
which is proportional to the size of the search space (that is inefficient).

\section{An Improvement of Network\label{sec:4}}
\begin{figure*}[t]
    \begin{center}
        \subfloat[Integer function $\tau_j(step)$]{
            \includegraphics[scale=0.25]{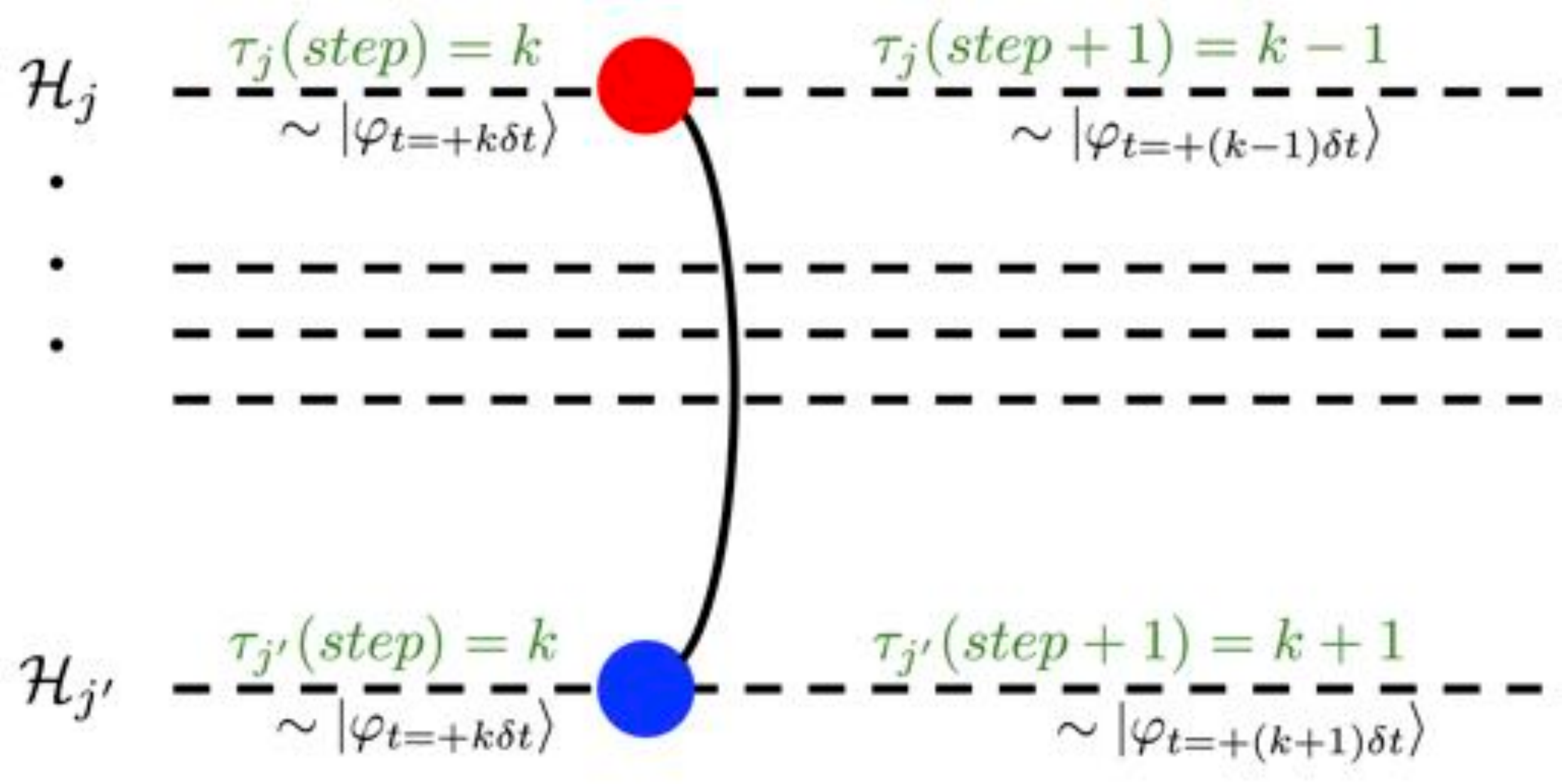}
        }~
        \subfloat[Example of the network with $m=4$.]{
            \includegraphics[scale=0.35]{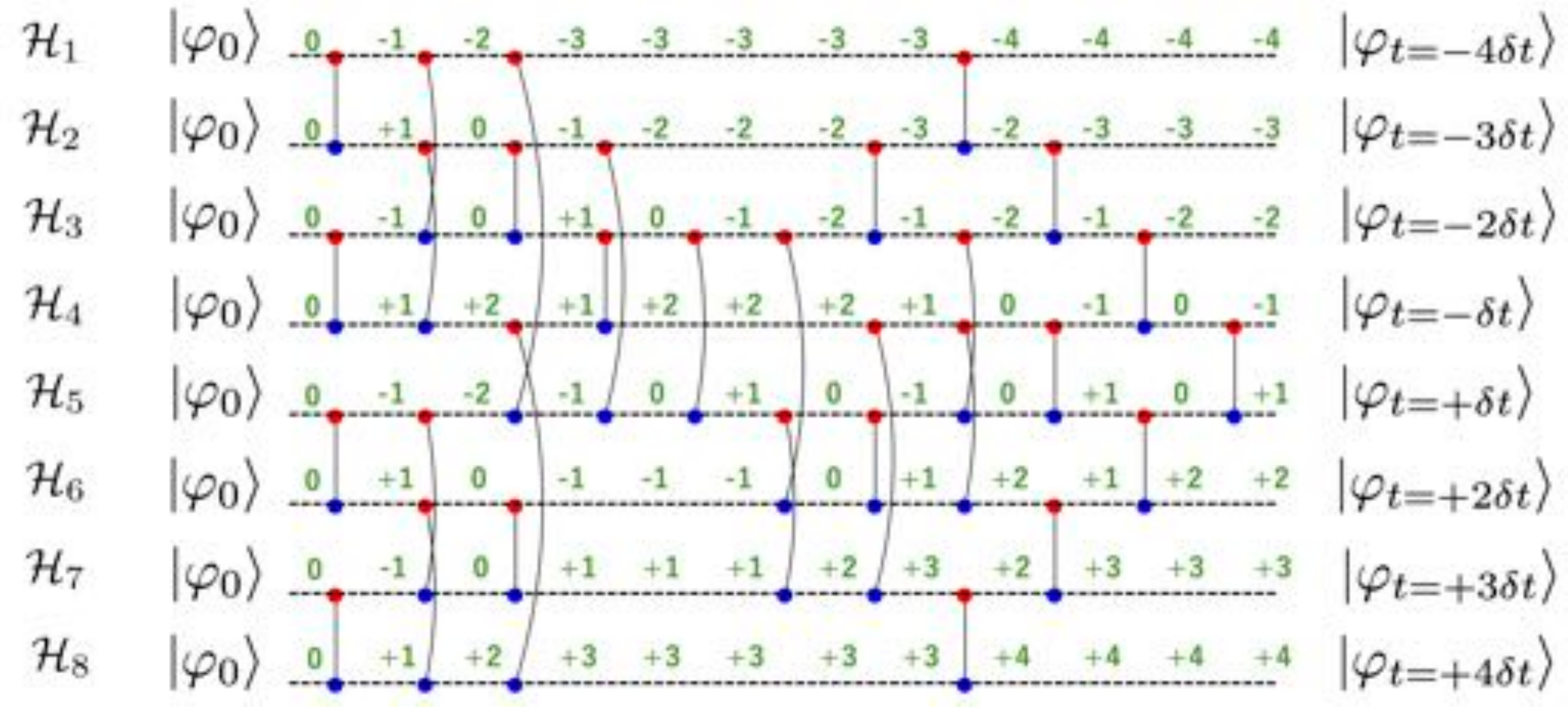}
        }\\
         \subfloat[Example of the network with $m=16$.]{
            \includegraphics[scale=0.5]{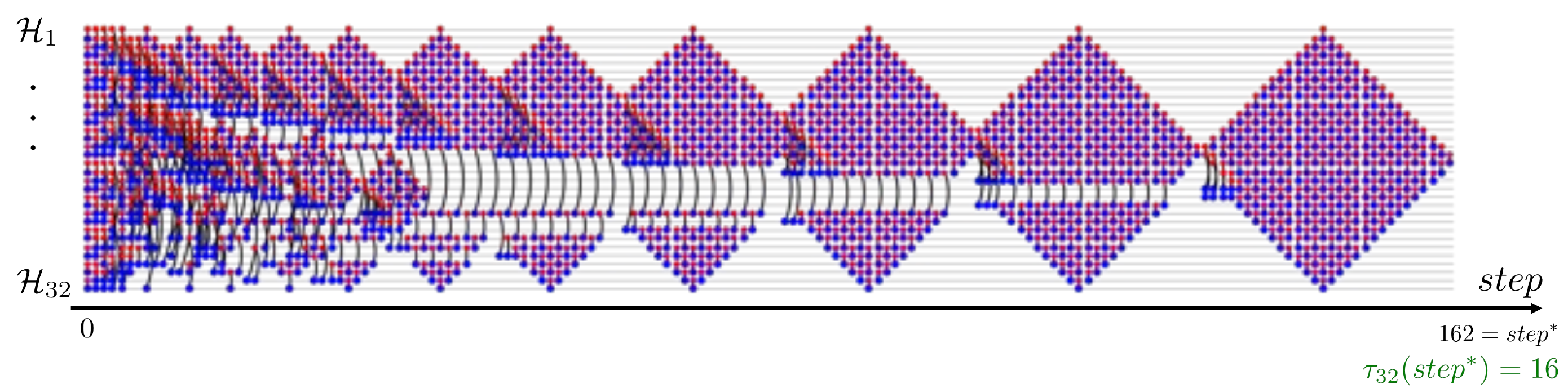}
        }
        \caption{{\bf Improved Network.} \label{fig:4}}
    \end{center}    
\end{figure*}

To resolve the inefficiency by the tournament network, we consider an improvement of the network for the emulation.
In the following, we make use of the fact that $|\varphi_{t=+(k+1) \delta t}\rangle$ and $|\varphi_{t=+(k-1) \delta t}\rangle$ are simultaneously obtained when the protocol is applied to two systems in $|\varphi_{t=+k\delta t}\rangle\otimes|\varphi_{t=+k\delta t}\rangle$. We construct a new network among Hilbert space $\mathcal{H}_{1}\otimes,\cdots,\otimes\mathcal{H}_{2m}$
$\left(=\mathcal{H}^{\otimes 2m}\right)$ as follows:\\
\begin{enumerate}
\item Introducing integer $step\in \{0,\cdots,step^*\}$, an integer function $\tau_j(step)$ is assigned to each Hilbert space $\mathcal{H}_j$. 
We put $\tau_j(0)=0$ is for all $j\in\{1,\cdots,2m\}$.
\item For a given $\{\tau_j(step)\}_{j\in\{1,\cdots, 2m\}}$, $\{\tau_j(step+1)\}_{j\in \{1,\cdots, 2m\}}$ is iteratively defined as follows:
\begin{enumerate}
\item Start with $j=1$.
\item If there is no $j'(>j)$ such as $\tau_j(step)=\tau_{j'}(step)$, define $\tau_{j}(step+1)$ by $$\tau_{j}(step+1):=\tau_{j}(step).$$
\item Otherwise, create a pair $\{j,j'\}$ with the minimum $j'$ that holds $\tau_{j}(step)=\tau_{j'}(step)$, define $\tau_{j}(step+1)$ and $\tau_{j'}(step+1)$ by
$$\tau_{j}(step+1):=\tau_j(step)-1$$
and
$$\tau_{j'}(step+1):=\tau_{j'}(step)+1.$$
\item Add the pair $\{j,j'\}$ as an element of the set $\Sigma(step)$.
\item Set $j=j''$ where $j'' (>j)$ is the minimum integer that does not appear yet in any pair included in $\Sigma(step)$, go back to (b) until $j=2m$.
\end{enumerate}
\item Making increment as $step=step+1$, go back to 2. \\ (Repeat the increment, unless $\Sigma(step)=\emptyset$.) 
\item Let $step^*$ be the final value of $step$.
\end{enumerate}
(See also FIG.\ref{fig:4}.)
We can numerically verify the existence of such $step^*$ and $step^*=O(m^2)$. In each $step\in \{1,\cdots,steps^*\}$, we apply our protocol to the pairs of two systems on ${\mathcal H}_j$ and ${\mathcal H}_{j'}$ appearing in $\Sigma(step)$. Suppose that $\rho_j(step)\otimes \rho_{j'}(step)$ exists on ${\mathcal H}_j\otimes{\mathcal H}_{j'}$ where each state is given
\begin{eqnarray}
\rho_j(step)&&=|\varphi_{t=\tau(step)\delta t}\rangle\langle\varphi_{t=\tau(step)\delta t}|\nonumber\\
&&+\delta t^2 \delta\rho_j+O(\delta t^3)
\label{eq:34}
\end{eqnarray}
and
\begin{eqnarray}
\rho_{j'}(step)&&=|\varphi_{t=\tau(step)\delta t}\rangle\langle\varphi_{t=\tau(step)\delta t}|\nonumber\\
&&+\delta t^2 \delta\rho_{j'}+O(\delta t^3)
\label{eq:35}
\end{eqnarray}
respectively. Being applied the protocol, the states are evolved as
\begin{eqnarray}
\rho_{j}(step)&&\mapsto
|\varphi_{t=(\tau(step)-1)\delta t}\rangle\langle\varphi_{t=(\tau(step)-1)\delta t}|\nonumber\\
&&
\!\!\!\!\!\!
-\delta t^2\Big(\left\{\hat{\Gamma}_{\tau(step)\delta t}, |\varphi_{t=\tau(step)\delta t}\rangle\langle\varphi_{t=\tau(step)\delta t}|\right\}\nonumber\\
&&-2 \langle\varphi_{t=\tau(step)\delta t}|\hat{\Gamma}_{\tau(step)\delta t}|\varphi_{t=\tau(step)\delta t}\rangle\Big)
\nonumber\\
&&+\frac{\delta t^2}{2} 
\left(\delta\rho_j+\delta\rho_{j'}\right)+O(\delta t^3)
\end{eqnarray}
and
\begin{eqnarray}
\rho_{j'}(step)&&\mapsto
|\varphi_{t=(\tau(step)+1)\delta t}\rangle\langle\varphi_{t=(\tau(step)+1)\delta t}|\nonumber\\
&&
\!\!\!\!\!\!
-\delta t^2\Big(\left\{\hat{\Gamma}_{\tau(step)\delta t}, |\varphi_{t=\tau(step)\delta t}\rangle\langle\varphi_{t=\tau(step)\delta t}|\right\}\nonumber\\
&&-2 \langle\varphi_{t=\tau(step)\delta t}|\hat{\Gamma}_{\tau(step)\delta t}|\varphi_{t=\tau(step)\delta t}\rangle\Big)
\nonumber\\
&&+\frac{\delta t^2}{2} 
\left(\delta\rho_j+\delta\rho_{j'}\right)+O(\delta t^3).
\end{eqnarray}
Notice that, undergoing each protocol, the deviation (from the first term) which is order of $O(\delta t^2)$ will be additionally accumulated.
In order to reduce such deviations as much as possible, we employ an additional procedure replacing the states in (\ref{eq:34}) and (\ref{eq:35}) by the fresh initial state $|\varphi_{t=0}\rangle\langle\varphi_{t=0}|$ for the pair with $\tau(step)=0$ and $step\not=0$.

For the given the network, starting from initial state $|\varphi_{t=0}\rangle^{\otimes 2m} \in {\mathcal H}^{\otimes 2m}$ with $step=0$, being applied the above repeatedly to the pairs indicated by the network, the state in Hilbert space ${\mathcal H}_j$ finally becomes 
\begin{eqnarray}
\rho_{j}(step^*)&&=|\varphi_{t=\tau_j(step^*)\delta t}\rangle\langle\varphi_{t=\tau_j(step^*)\delta t}|\nonumber\\
&&-\delta t^2 \sum_{k=1}^{2m+1} K_{j,k}^{(2m)}
\left(\{\hat{\Gamma}_{k'},|\varphi_{t=k'\delta t}\rangle\langle\varphi_{t=k'\delta t}|\}\right.\nonumber\\
&&~~~~~~\left.-2\langle\varphi_{t=k'\delta t}|\hat{\Gamma}_{k'}|\varphi_{t=k'\delta t}\rangle\right)+O(\delta t^3)
\label{eq:39}
\end{eqnarray}
where $k'=k-m-1$ and $K_{j,k}^{(2m)}$ is the consequence of the above mentioned accumulation of the deviation from the first term.  Notice that the coefficient can be determined only by the structure of the network but independently from Hamiltonian $\hat{H}$. (Further properties of the coefficient will be addressed shortly after.) By the construction of the network, $\tau_j(step^*)$ is uniquely determined as
\begin{equation}\label{eq:40}
\tau_j(step^*)=\left\{
\begin{array}{ll}
j-m-1 & \mbox{for}~1\le j\le m\\
j-m & \mbox{for}~m< j \le 2m.
\end{array}\right.
\end{equation}
Thus, the state in Hilbert space ${\mathcal H}_{2m}$ particularly becomes
\begin{eqnarray}
\rho_{2m}(step^*)&&=|\varphi_{t=+m\delta t}\rangle\langle\varphi_{t=+m\delta t}|\nonumber\\
&&-\delta t^2 \sum_{k=1}^{2m+1} K_{2m,k}^{(2m)}
\left(\{\hat{\Gamma}_{k'},|\varphi_{t=k'\delta t}\rangle\langle\varphi_{t=k'\delta t}|\}\right.\nonumber\\
&&~~~\left.-2\langle\varphi_{t=k'\delta t}|\hat{\Gamma}_{k'}|\varphi_{t=k'\delta t}\rangle\right)+O(\delta t^3).\label{eq:41}
\end{eqnarray}
The required number of the systems to make the first term achieve $|\varphi_{t=t_c}\rangle\langle\varphi_{t=t_c}|$ is given as
\begin{equation}\label{eq:42}
2m\sim O(\log\dim({\mathcal H}))
\end{equation}
that seems an exponential improvement of (\ref{eq:34-1}). To make the improvement veridical, we need to check that the second term in (\ref{eq:41}) is subdominant in comparison with the first term. Because of the existence of coefficient $K_{j,k}^{(2m)}$, the estimation of the second term becomes nontrivial in comparison with the case of the tournament network. We numerically find that the coefficient approximately follows a scaling law as
\begin{equation}\label{eq:43}
K_{j,k}^{(2m)}\sim \Lambda^{-1} K_{\Lambda j,\Lambda k}^{(2 \Lambda m)}
\end{equation} 
with a positive real parameter $\Lambda$. (See appendix too for the scaling law.) That scaling law implies $K_{j,k}^{(2m)}=O(m)$. Together with $m=O(\delta t^{-1})$, the order of the coefficient can be estimated as $K_{j,k}^{(2m)}=O(\delta t^{-1})$, and the order of the second term of (\ref{eq:41}) including $\delta t^2$-factor can be estimated as
\begin{eqnarray}
\hat{\Xi}_{2m}:=&&
\delta t^2 \sum_{k=1}^{2m+1} K_{2m,k}^{(2m)}
\left(\{\hat{\Gamma}_{k'},|\varphi_{t=k'\delta t}\rangle\langle\varphi_{t=k'\delta t}|\}\right.\nonumber\\
&&\left.-2\langle\varphi_{t=k'\delta t}|\hat{\Gamma}_{k'}|\varphi_{t=k'\delta t}\rangle\right)\nonumber\\
&&=O(\delta t^0).
\label{eq:44}
\end{eqnarray}
\begin{figure*}[t]
    \begin{center}
        \subfloat[]{
            \includegraphics[scale=0.5]{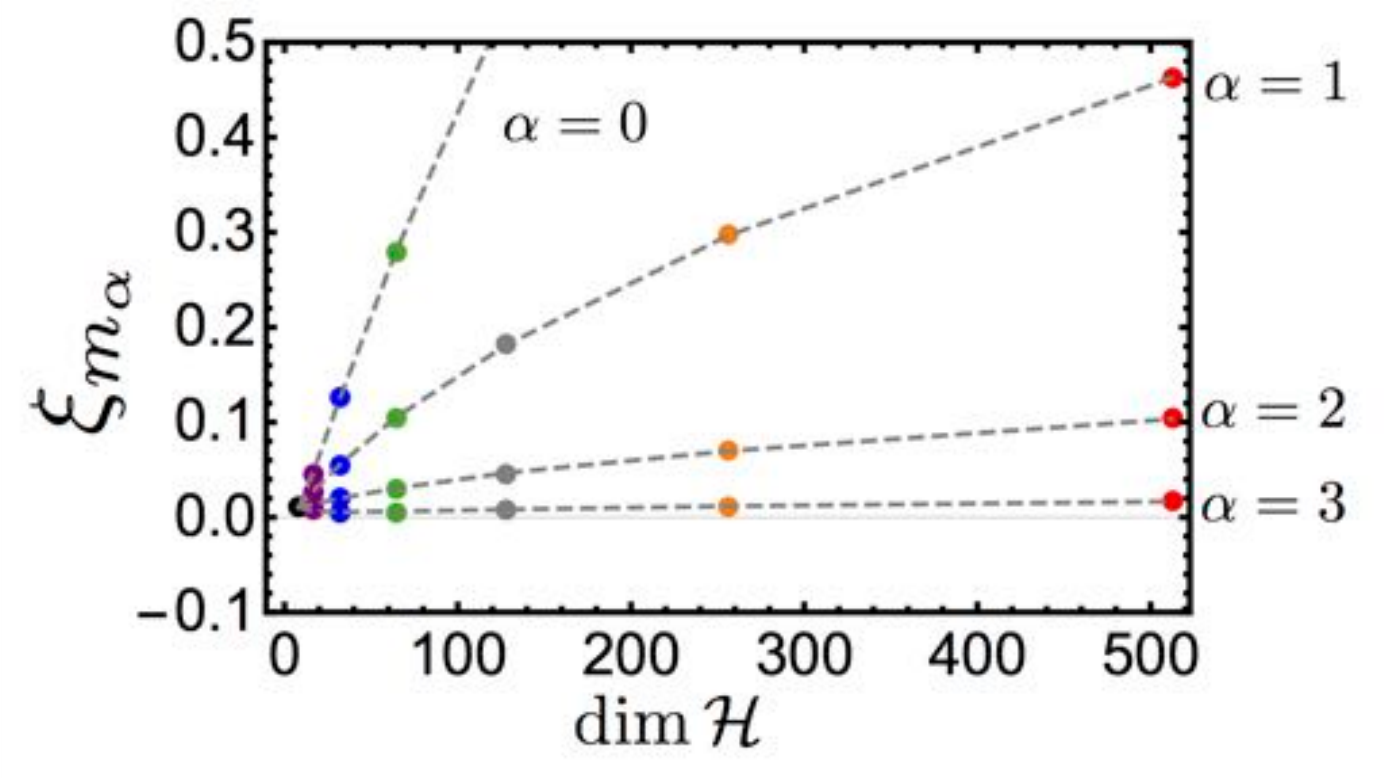}
        }
        \subfloat[]{
		\includegraphics[scale=0.5]{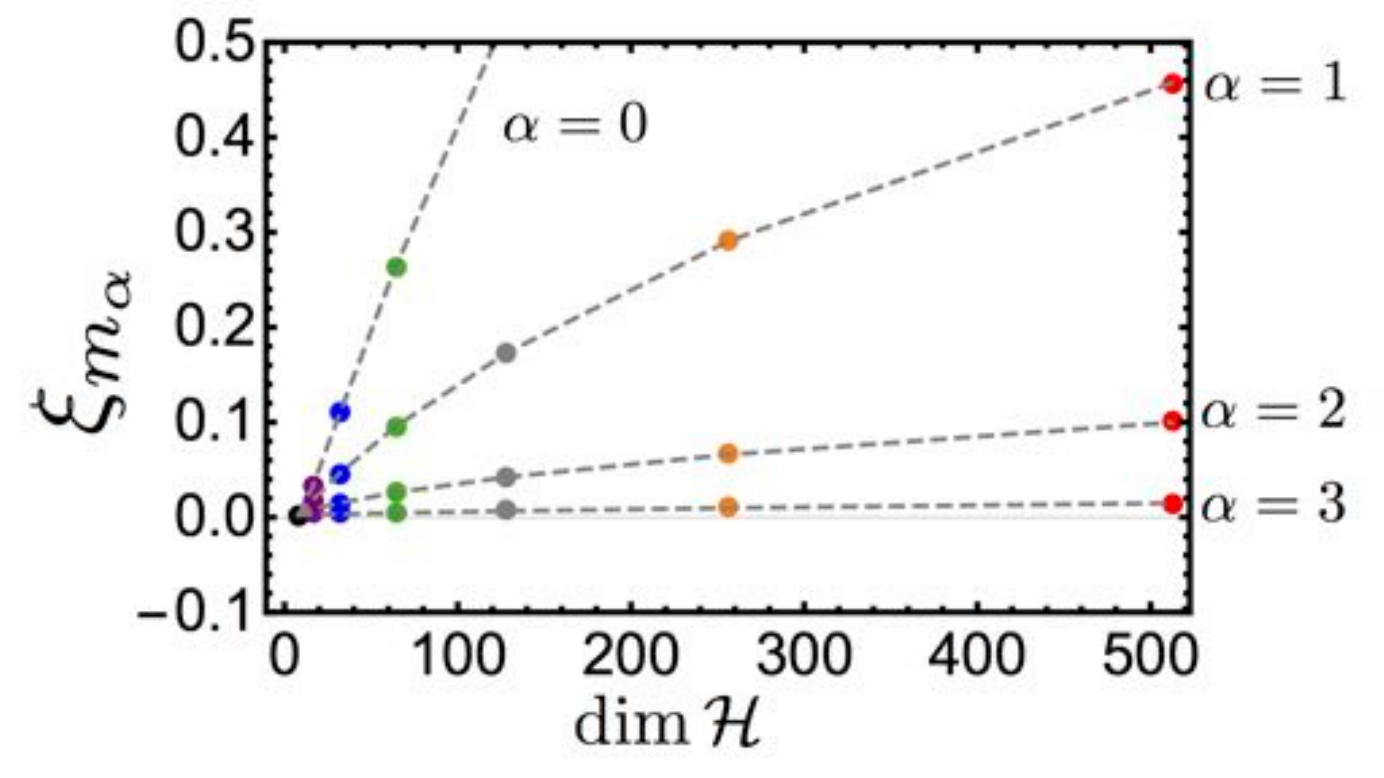}
        }\\
        \subfloat[]{
            \includegraphics[scale=0.5]{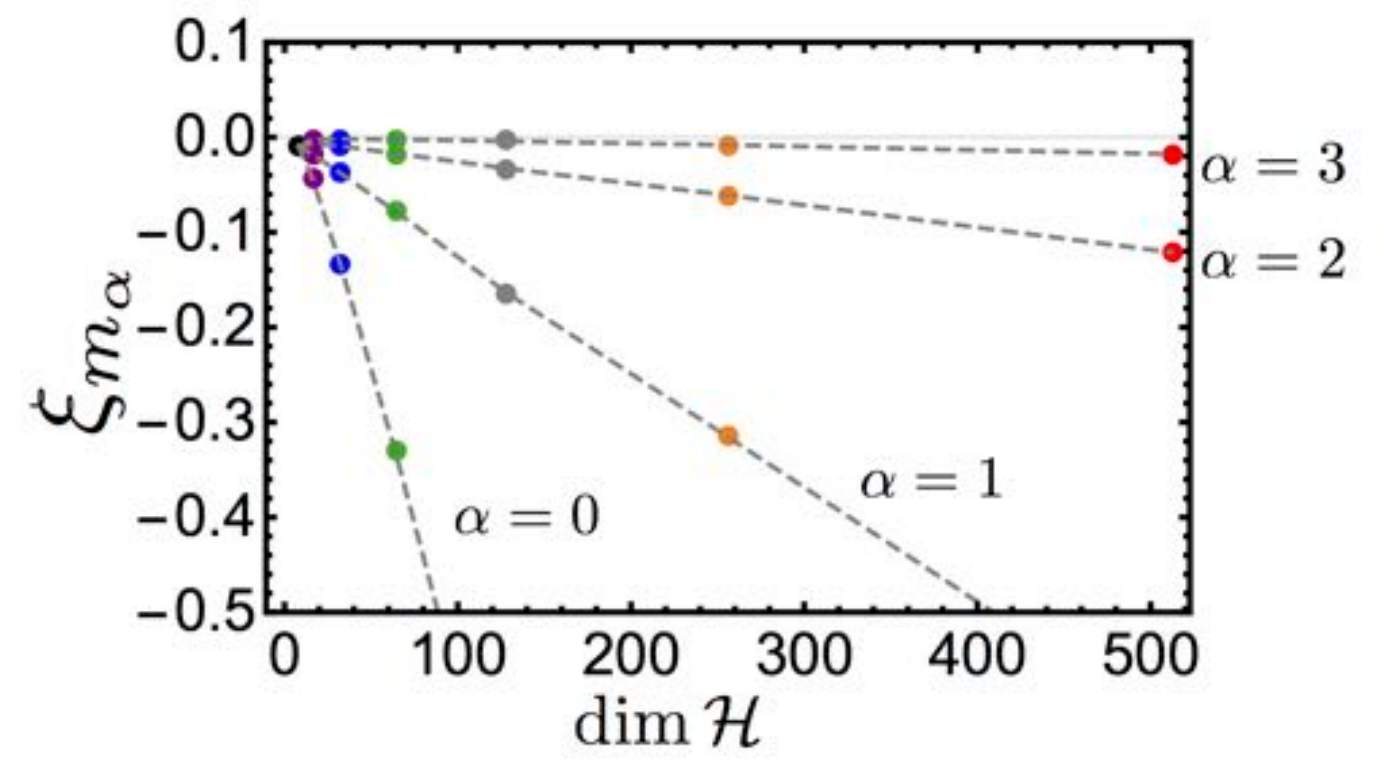}
        }
        \subfloat[]{
            \includegraphics[scale=0.5]{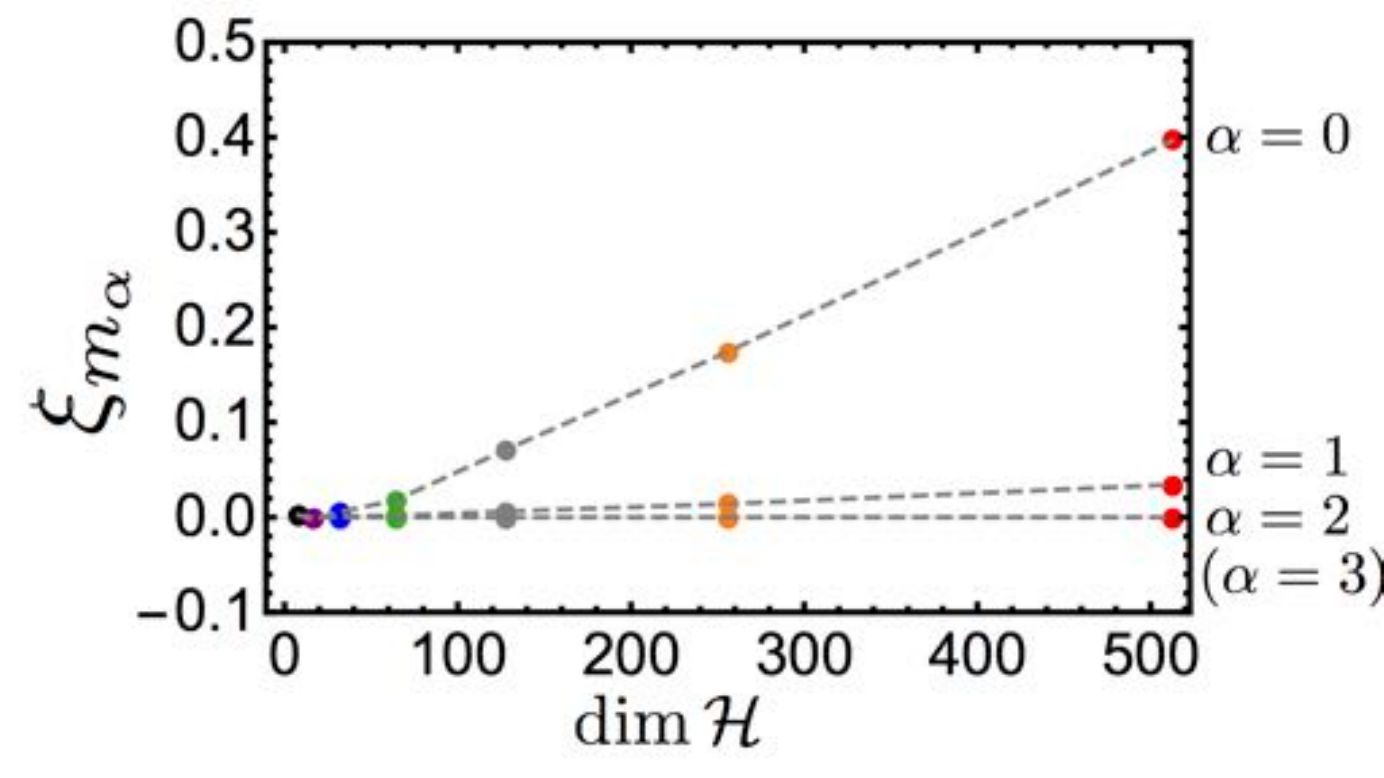}
        }
        \caption{{\bf Plot of $\xi_{m_\alpha}$ in (\ref{eq:46}).} All numerical computations are done with $\delta t=0.01 \Delta^{-1}$. Each $K^{(2m_\alpha)}_{j,k}$ used in the computation is obtained from $K^{(2\times 128)}_{j,k}$ following the scaling low in (\ref{eq:43}). (Since the case in (a) is contradicting to the necessary condition in (\ref{ineq2}), the plot of (a) is shown merely as a reference.)\label{fig:5}}
    \end{center}    
\end{figure*}
Unlike (\ref{eq:33}), we cannot control the order of the magnitude of the second term by choosing a smaller $\delta t$. In other words, for such $m$, the reason the higher order terms to be less dominant cannot rely on the order of $\delta t$ itself. Further careful estimation is required to check if the second term is still subdominant even under such situation. We will numerically address this issue in the following section.

\section{Remarks on the Improved Network\label{sec:5}}
Besides the issue of the term in (\ref{eq:44}), we have the following fundamental constraint that may hinder (\ref{eq:42}) to be veridical: Remember that the protocol we proposed in SEC.\ref{sec:2} preserves the total energy on the two systems as is clearly shown in (\ref{eq:10}) and (\ref{eq:11}). In other words, the total energy of the $2m$ systems participating to the proposed network must be preserved. If the first term in (\ref{eq:41}) is dominant, the total energy of the $2m$ systems is given as
\begin{equation}
\sum_{k=1}^{2m} \langle\varphi_{t=k\delta t}|\hat{H}|\varphi_{t=k\delta t}\rangle\simeq 2m E_0
\end{equation}
From the above, we obtain
\begin{equation}\label{ineq}
m >\frac{1}{2}\frac{\varepsilon_{\dim({\mathcal H})}-\varepsilon_{1}}{\varepsilon_{\dim({\mathcal H})}-E_0}.
\end{equation}
With examples (b), (c), and (d) in TABLE \ref{table:table1}, since $E_0=0$ and $\varepsilon_1=-\varepsilon_{\dim{(\mathcal{H})}}$ hold, the above inequality does not give any practical condition for $m$. On the contrary, however,
since $E_0=-\Delta/\dim({\mathcal H})$ and $\varepsilon_1=0$ hold with example (a),
\begin{equation}\label{ineq2}
m > \frac{\dim({\mathcal H})}{2}
\end{equation}
is implied. The bound is obviously contradicting to (\ref{eq:42}), and the network would not work as is expected. In other words, there certainly exists a necessary condition to make the network appropriately work in the spectrum structure of the Hamiltonian. 
The necessary condition, however, can be always satisfied by employing the following trick. For a given Hamiltonian $\hat{H}$ on ${\mathcal H}$, introducing a doubling Hilbert space $\tilde{\mathcal H}={\mathcal H}\otimes {\mathcal H}$, we can always introduce a doubled Hamiltonian
\begin{equation}
\tilde{H}=\hat{H}\otimes I- I \otimes \hat{H}
\end{equation}
instead of the original $\hat{H}$. Notice that the ground state of $\tilde{H}$ is $|g\rangle\otimes |e\rangle$ where $|g\rangle$ and $|e\rangle$ is the ground and the most excited state of $\hat{H}$ respectively. (Similarly, the most excited state of $\tilde{H}$ is $|e\rangle\otimes|g\rangle$.) Thus, if we succeed to dynamically search the ground state of $\tilde{H}$, we can obtain the ground state of $\hat{H}$ at the same time. Moreover, when we apply our approach to $\tilde{H}$ instead of $\hat{H}$, since $E_0=0$ and $\varepsilon_{\dim(\tilde{\mathcal{H}})}=-\varepsilon_{1}$ always hold, the inequality in (\ref{ineq}) does not give any practical condition for $m$. Thus, if necessary, we can always employ this doubling trick to resolve the condition in (\ref{ineq}). (Example (b) in TABLE\ref{table:table1} with $\dim({\mathcal H})=N^2$ corresponds to the doubling trick applied to example (a) with $\dim({\mathcal H})=N$.)

Now, let us go back to the estimation of (\ref{eq:44}). As a simplest trial, we numerically compute 
\begin{equation}\label{eq:46}
\xi_{m_\alpha}:=
\langle\varphi_{t=+m_{\alpha} \delta t}|~\hat{\Xi}_{2m_{\alpha}}|\varphi_{t=+m_{\alpha}\delta t}\rangle
\end{equation}
where $m_{\alpha}$ is the minimum $m$ which holds
\begin{equation}\label{eq:47}
P_1(+m\delta t)=\frac{1}{2}\left(\frac{\log_2 8}{\log_2 \dim{\mathcal H}}\right)^{\alpha},
\end{equation}
with integer $\alpha$. Condition $|\xi_{m_\alpha}| \ll 1$ would be a necessary condition to be sure that the improved network appropriately works.
(Remember that $P_1(0)=(\dim{\mathcal H})^{-1}$. Probability $P_1(+m_\alpha\delta t)$ is exponentially large in comparison with $P_1(0)$. We employ the particular form of (\ref{eq:47}) so that $m_{\alpha}$ is independent from $\alpha$ for $\dim{\mathcal H}=8$.) As is shown in FIG.\ref{fig:5}, we find that increasing of $\alpha$ restrains the amplitude of $\xi_{m_\alpha}$ with each example. More significantly, the amplitude of $\xi_{m^\alpha}$ tends to be saturated as $\dim{\mathcal H}$ increases. If that is the case, the behavior suggests that the proposed network with the protocol would work efficiently (i.e., within complexity of (polynomial of) the logarithmic with respect to $\dim {\mathcal H}$) to find the ground state of the Hamiltonian. Notice that  the Hamiltonian of example (a) (or example (b) as the doubled version of example (a)) corresponds to the database searching problem and that  the above suggestion might imply an exponential improvement of the well known results by Grover's algorithm \cite{PhysRevLett.79.325,Grover1996AFQ} or by the ordinary quantum annealing approach \cite{2000quant.ph..1106F,Farhi2000ANS,PhysRevA.65.042308}. Unfortunately, to be sure of the suggestion rigorously, we need further investigations beyond the numerical examination as remains to be our future challenge.

Concerning the number of the systems required in running the above, we can reduce the number by decreasing of the success probability of the finding the ground state. Instead of $2m$ systems in Hilbert space ${\mathcal H}^{\otimes 2m}$, let us consider a compound system described in Hilbert space ${\mathcal H}\otimes{\mathcal K}$ where $\dim{\mathcal K}=2m$. On the compound system, preparing an initial state
\begin{equation}
|\varphi_0\rangle\otimes \frac{1}{\sqrt{\dim{\mathcal K}}}\sum_{j=1}^{\dim{\mathcal K}} |j\rangle
\end{equation}
with orthonormal basis $\{|j\rangle\}_{j=1}^{\dim{\mathcal K}}$ of ${\mathcal K}$, we can apply our energy transfer protocol not to two systems in Hilbert spaces ${\mathcal H}_j$ and ${\mathcal H_{j'}}$ but to the two components of the state vector in the sectors spanned by $|j\rangle$ and $|j'\rangle$. Finally, the component of the state vector spanned by $|2m\rangle$ approximately achieves $|\varphi_{t=+m\delta t}\rangle$ in ${\mathcal H}$. Notice that the probability to obtain the component by measurement  is 
$(\dim{\mathcal K})^{-1}=(2m)^{-1}$, and that the probability is still logarithmic polynomial in $\dim{\mathcal H}$. Thus, even taking the probability into account, we have a good chance to find the ground state with a probability of logarithmic polynomial with respect to $\dim{\mathcal H}$. The system for Hilbert space ${\mathcal K}$ is implementable as an ancillary system with $\log_2 (2m)(=O(\log\log\dim{\mathcal H}))$ qubits that is efficient at least in a theoretical sense.

\section{Summary}
In this article, we have introduced the following two idea; 1) Energy transfer protocol among two systems, and 2) Network structure to efficiently apply the protocol to the ground state search problem. Below we make some additional comments on each of them. 

First, let us remark on the conceptually interesting point of the first protocol. As is well known, Hamiltonian has two significant roles in the quantum mechanics in general; one is as an observable corresponding to energy, another is as a generator of the time evolution of the dynamics. As the consequence, the energy of the system is always conserved under the dynamics naturally generated by the Hamiltonian itself. To change the energy of the system dynamically, we need something else besides Hamiltonian. What we found here, on the other hand, is a use of the interference induced by the quantum swapping among two systems in forward and backward evolved state. The interference can create the one-way energy transfer among the two systems, while the total energy of the two systems is conserved.  In defining the protocol, we do not need any extras but only Hamiltonian itself, swapping operation, and time duration parameter $\delta t$.  The simple structure makes us imagine a good relations of the protocol to fundamental aspects of quantum mechanics. In fact,  (\ref{eq:13}) and (\ref{eq:14}) can be interpreted as  ``the time duration $\delta t\sim\delta E_0^{-1}$ is required to achieve the energy transfer of the magnitude of $\delta E_0$ that each system originally has". That seems a manifestation of the time-energy uncertainty relation that has not been well investigated before. Besides it, we expect that the protocol might give a new insight into a relation among the three fundamental topics in quantum mechanics, i.e., energy, time evolution, and interference.

Concerning the network part, we remain some challenges to be done.  First of all, although our numerical analysis seems positively suggest the existence of examples with which our approach works, further rigorous analysis to check if such suggestion can be proven or not will be indispensably required. The issue is left as future challenge of our approach. Besides it, {\it like all proposals for quantum computation, relies on speculative technology, does not in its current form take into account all possible sources of noise, unreliability and manufacturing error, and probably might not work \cite{Laudauer}}. For the reason, estimations of stability or fault tolerance of our approach against any imperfections would be another indispensable challenge we need to address. In addition, combinations with some quantum error correction technique would be also exciting challenge from both theoretical and practical points of view.

\begin{acknowledgments}
This work is based on results  obtained from a project commissioned by New Energy and Industrial Technology Development Organization(NEDO),Japan.
\end{acknowledgments}

\appendix
\section{Derivation of inequality (\ref{eq:30})}
From eqs. (\ref{eq:19}), (\ref{eq:20}) and (\ref{eq:21}), we obtain
\begin{equation}
\frac{d}{dt}P_1(t)=-\left(\varepsilon_1-\langle\varphi_t|\hat{H}|\varphi_t\rangle\right)P_1(t).
\end{equation}
Noticing 
$$
\varepsilon_1 P_1(t)+\varepsilon_2 \left(1-P_1(t)\right)\le \langle\varphi_t|\hat{H}|\varphi_t\rangle
$$
and
$$
\langle\varphi_t|\hat{H}|\varphi_t\rangle \le
\varepsilon_1 P_1(t)+\varepsilon_{\dim{\mathcal H}} \left(1-P_1(t)\right),
$$
we find that
\begin{equation}
\Delta ~P_1(t)\left(1-P_1(t)\right)\le \frac{d}{dt}P_1(t)
\end{equation}
and
\begin{equation}
\frac{d}{dt}P_1(t)\le
D~P_1(t)\left(1-P_1(t)\right)
\end{equation}
hold. Notice that, when the equality holds, the differential equation for $P_1(t)$ is so-called the logistic equation \cite{Verhulst1845,Wolfram2002}  whose analytical solution is known. With the initial state in (\ref{eq:27}), we obtain (\ref{eq:30}).

\section{On the scaling law in (\ref{eq:43})}
\begin{figure*}[t]
\centering
        \subfloat[$m=16$]{
            \includegraphics[scale=0.35]{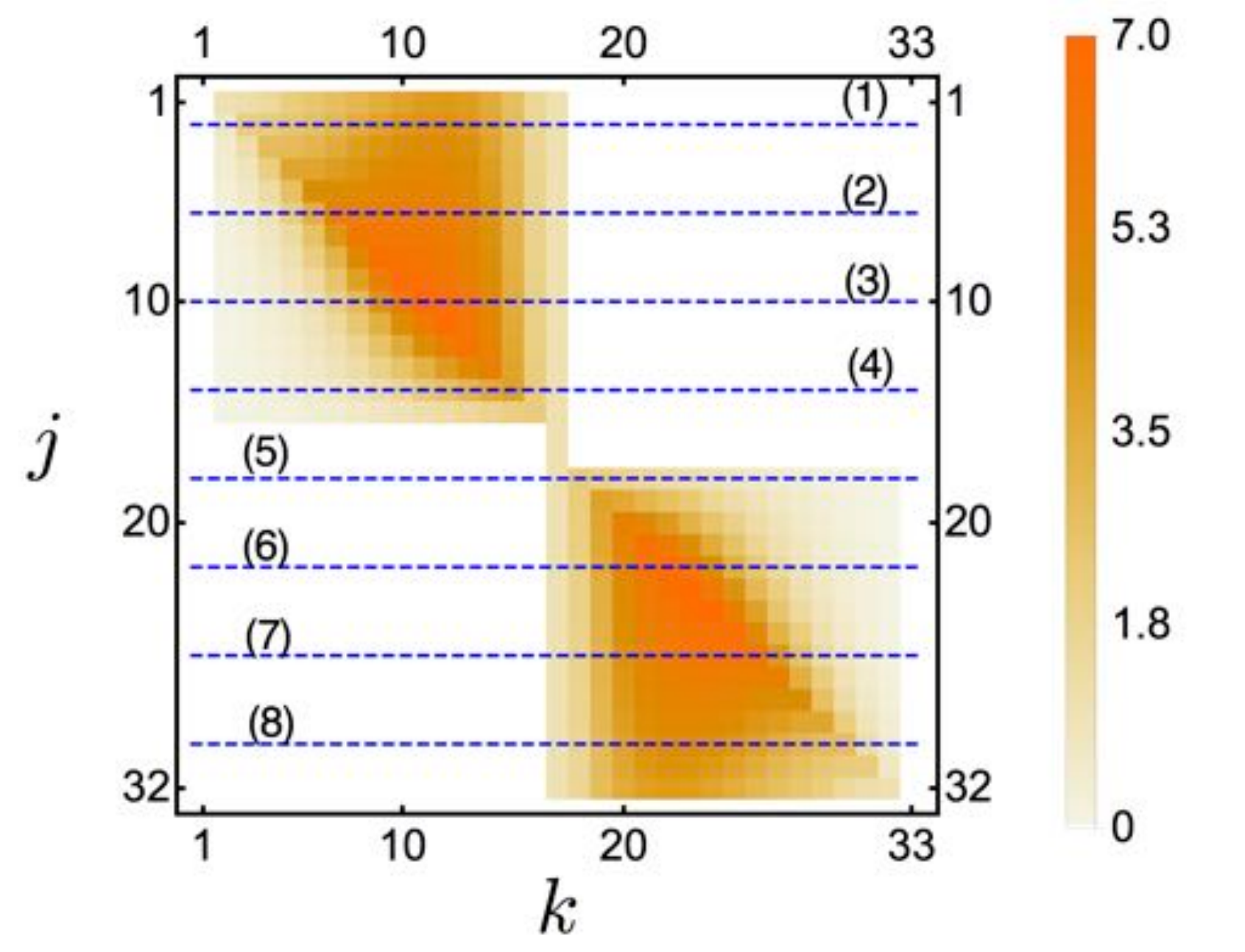}
        }~~~
        \subfloat[$m=32$]{
            \includegraphics[scale=0.35]{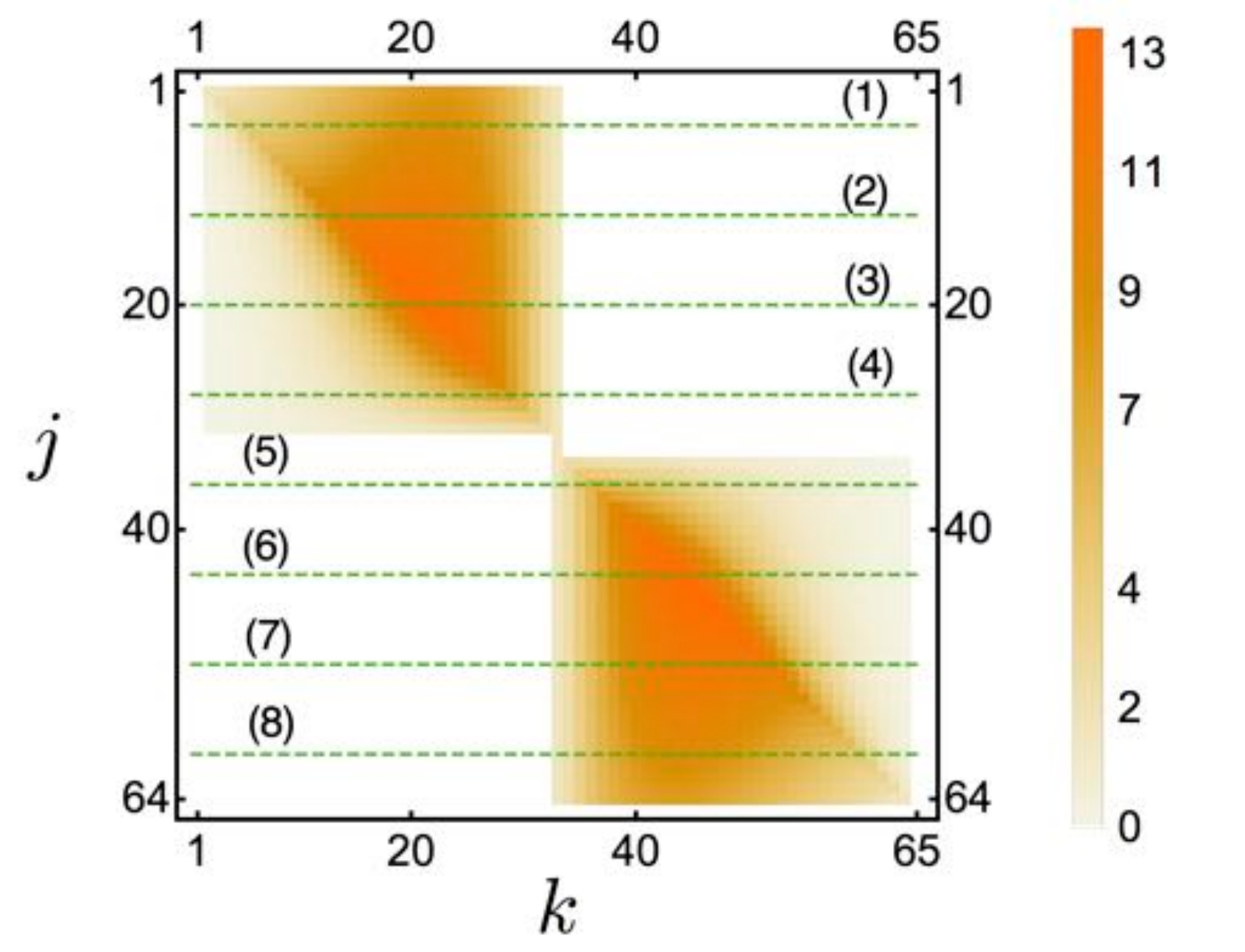}
        }\\
         \subfloat[$m=64$]{
            \includegraphics[scale=0.35]{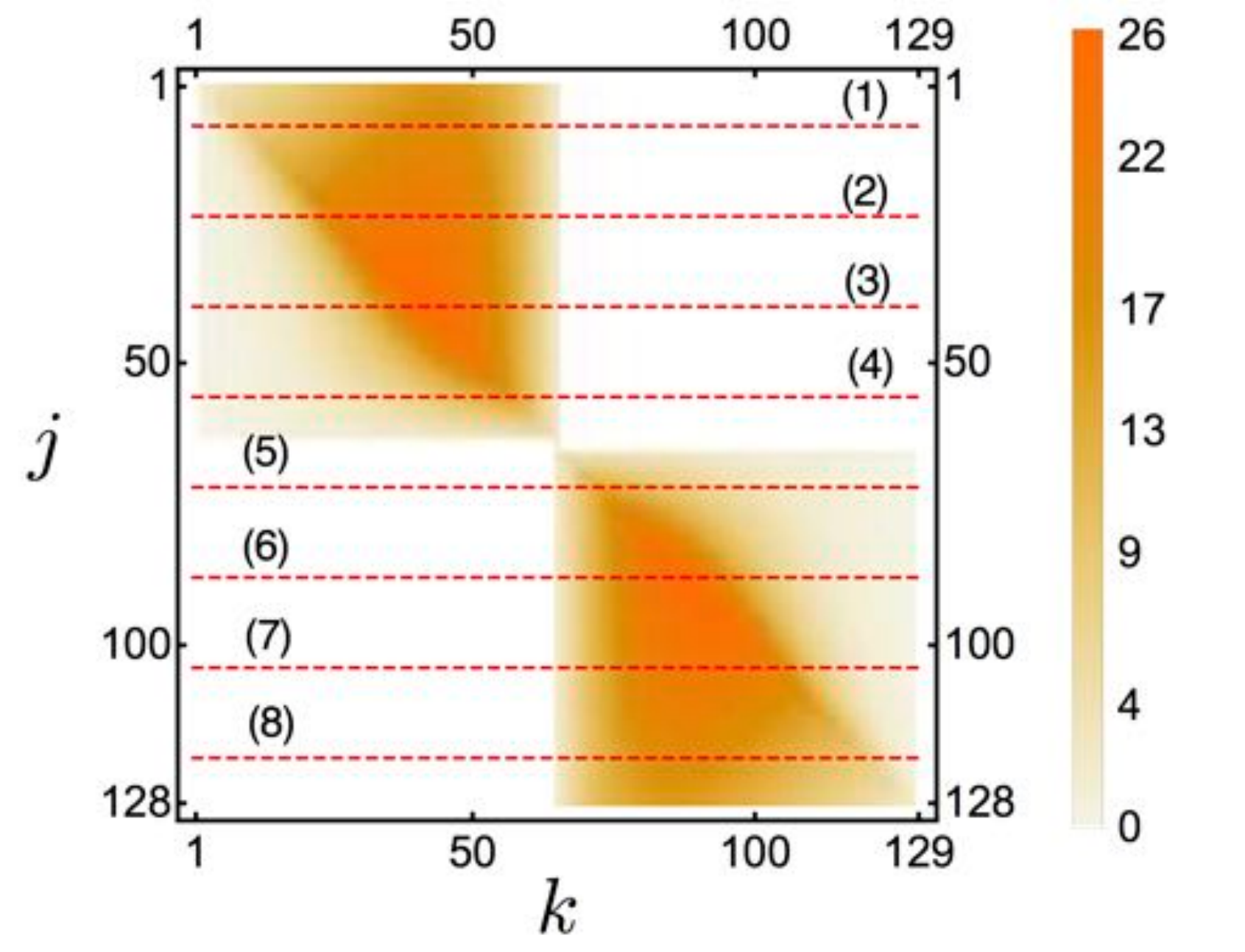}
        }~~~
        \subfloat[$m=128$]{
            \includegraphics[scale=0.35]{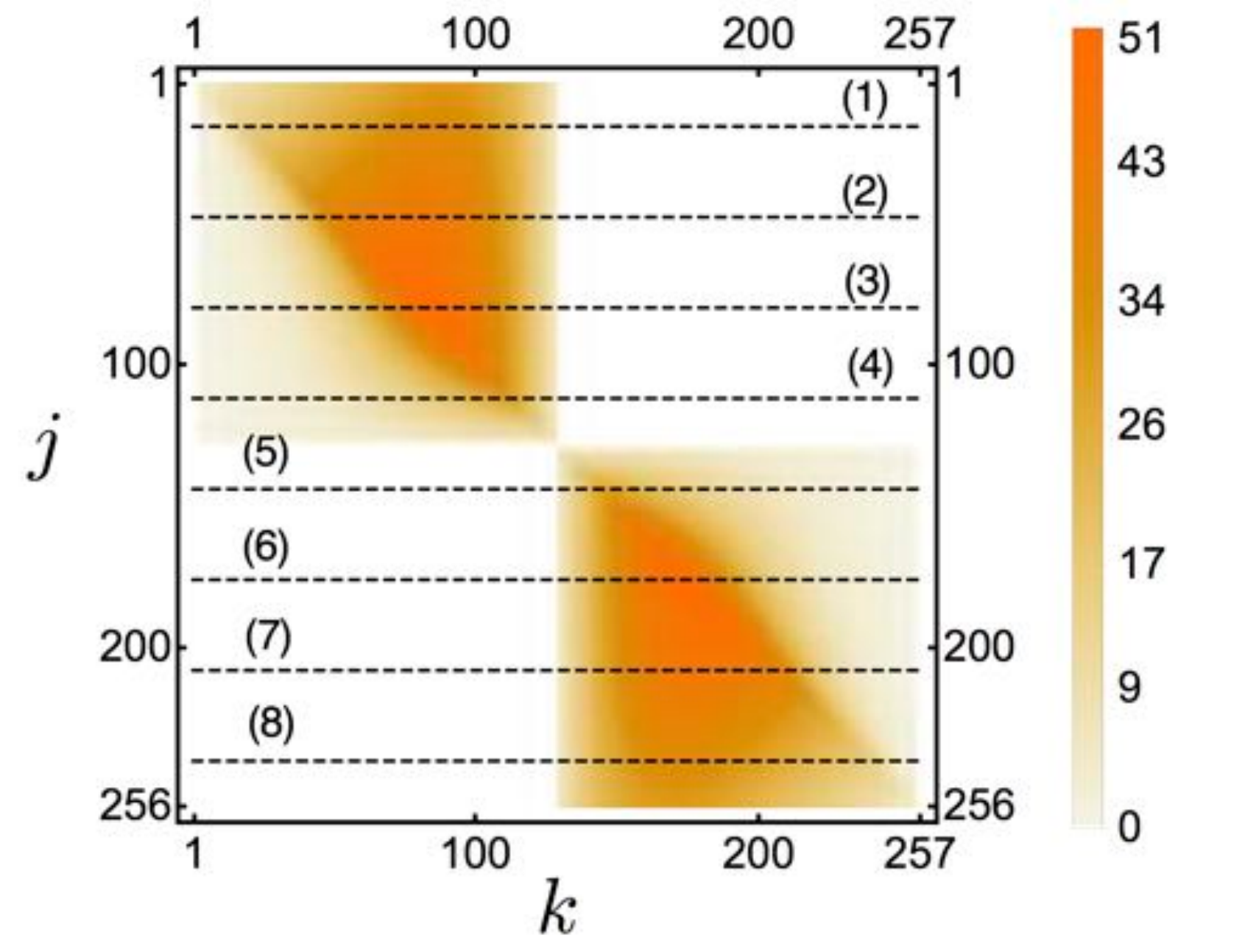}
        }\\
        \subfloat[Cross sections rescaled by (\ref{eq:43})]{
            \includegraphics[scale=0.45]{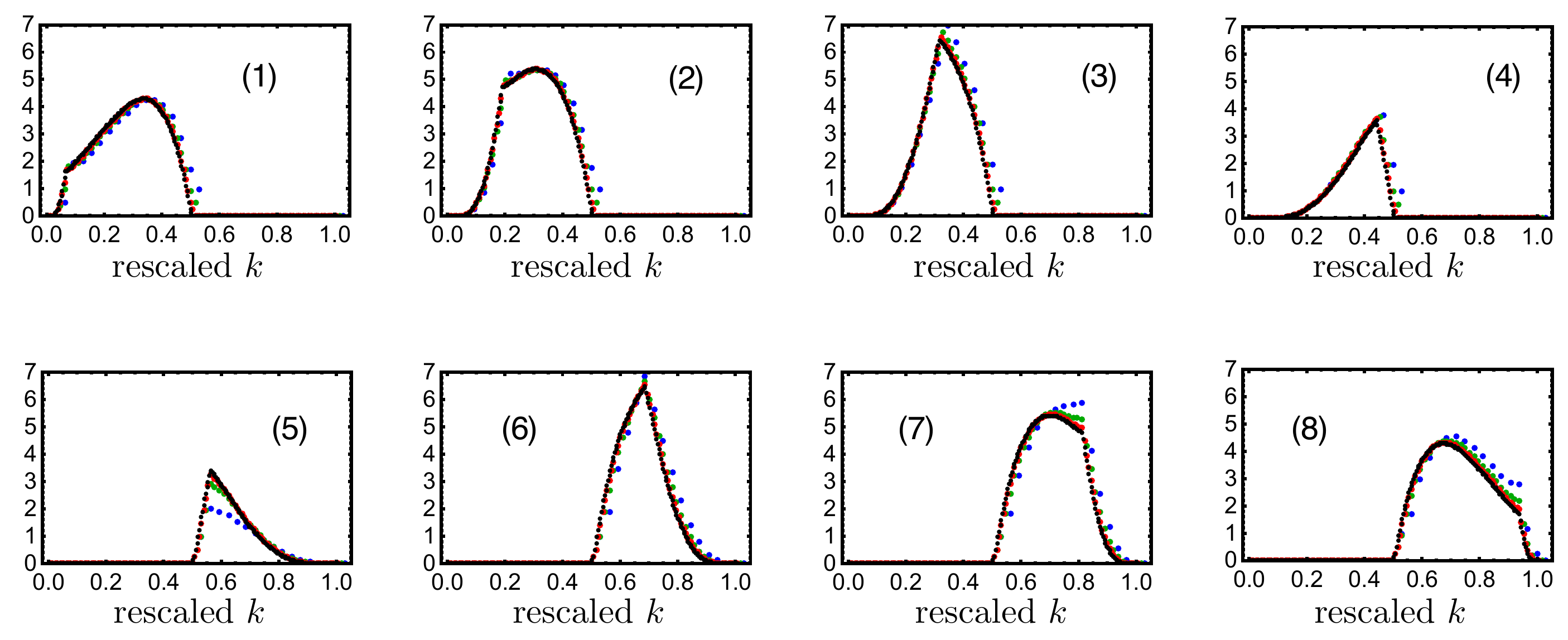}
        }
\caption{{\bf On the scaling law in (\ref{eq:43})}. In (a)--(d), the numerically computed amplitude of $K_{j,k}^{(2m)}$ is plotted with $m=16, 32, 64$ and $128$ respectively. Each of the eight panels in (e) corresponds to the dashed line (1)--(8) in (a)--(d). The plot colors (blue, green, red and black) in (e) correspond to (a), (b), (c) and (d) respectively. In (e), the amplitude are rescaled following  (\ref{eq:43}) based on the case of $m=16$.
\label{fig:afig1}}
\end{figure*}

In this section, we numerically show the scaling behavior of $K^{(2m)}_{j,k}$ described in (\ref{eq:43}). In (a)--(d) of Fig.\ref{fig:afig1}, we show the amplitude of $K^{(2m)}_{j,k}$ numerically computed with $m=16, 32, 64$ and $128$ respectively. Their cross sections at $8$ dashed lines indicated in (1)--(8) are rescaled according to (\ref{eq:43}) and plotted in (e). As we see, the scaling law in  (\ref{eq:43}) seems well-justified.

\newpage

\bibliography{sw2017}

\end{document}